\documentclass[structabstract]{aa}  
\usepackage{amstext,amsmath}
\usepackage{graphicx}

\usepackage{txfonts}

\usepackage{natbib}
\begin{document}
 \title{Three-Dimensional Temperature Mapping of Solar Photospheric Fine Structure Using \ion{Ca}{ii}~H Filtergrams}

   \author{V. M. J.  Henriques 
          \inst{1,2}     
          }

   \institute{Institute for Solar Physics, Royal Swedish Academy of Sciences, Albanova University Center, 106 91 Stockholm, Sweden
         \and
             Stockholm Observatory, Dept. of Astronomy, Stockholm University, Albanova University Center, 106 91 Stockholm, Sweden\\
                          \email{vasco@astro.su.se}}

  \date{Received September 6, 2012; accepted October 10, 2012}

  \abstract{The wings of the \ion{Ca}{ii}~H and K lines provide excellent  photospheric temperature diagnostics. At the Swedish 1-meter Solar Telescope (SST), the blue wing of \ion{Ca}{ii}~H is scanned with a narrowband interference filter mounted on a rotation stage. This provides up to 0\farcs10 spatial resolution filtergrams at high cadence that are concurrent with other diagnostics at longer wavelengths.}{The aim is to develop observational techniques that provide photospheric temperature stratification at the highest spatial resolution possible and use them to compare simulations and observations at different heights. }{We use filtergrams in the \ion{Ca}{ii}~H blue wing that were obtained with a tiltable interference filter at the SST. Synthetic observations are produced from three-dimensional (3D) hydro and magneto-hydrodynamic numerical simulations and degraded to match the observations. The temperature structure obtained from applying the method to the synthetic data is compared with the known structure in the simulated atmospheres and with observations of an active region. Cross-correlation techniques using restored non-simultaneous continuum images are used to reduce high-altitude, small-scale seeing signal introduced from the non-simultaneity of the frames when differentiating data.}{Temperature extraction using high-resolution filtergrams in the \ion{Ca}{ii}~H blue wing works reasonably well when tested with simulated 3D atmospheres. The cross-correlation technique successfully compensates for the problem of small-scale seeing differences and provides a measure of the spurious signal from this source in differentiated data. Synthesized data from the simulated atmospheres (including pores) match well the observations morphologically at different observed heights and in vertical temperature gradients.}{}
 

\authorrunning{V. M. J. Henriques}

\titlerunning{3D Temperature Mapping using the \ion{Ca}{ii}~H line.}

   \maketitle
%

\section{Introduction}

While the \ion{Ca}{ii}~H and K lines have been used for chromospheric studies for a long time, their usefulness as a photospheric temperature diagnostic remains largely unexplored in filtergrams with present-day large-aperture telescopes.

The extended damping wings of \ion{Ca}{ii}~H and K are insensitive to line-of-sight velocities and magnetic fields. They provide information over a wide depth range, are formed in Local Thermal Equilibrium (LTE), and allow data to be recorded at close to the highest spatial resolution achievable with ground-based optical telescopes. The absence of Doppler signals is an advantage when extracting temperature information. Furthermore, the upper photosphere processes that are sampled in the inner wings of these lines provide important clues to studying the chromosphere. \cite{2008A&A...479..213B} made a good case for this and used spectra in the inner \ion{Ca}{ii}~H blue wing to make a comparison with theoretical heating models and to identify photospheric waves as a driver for chromospheric line-core emission. 

Slit spectroscopy in the wings of these lines has been used to successfully constrain semi-empirical one-dimensional model atmospheres for the ``average Sun'' \citep{1975ApJ...201..799A,1976ApJ...205..874A,1987ApJ...312..909W} as well as for small magnetic feature models  \citep{2001ScChA..44..528F}. Also targeting bright magnetic features, \cite{1974SoPh...37..145S} used the \ion{Ca}{ii}~K line wings, the fact that \ion{Ca}{ii} is mostly in the ground state throughout the formation height of the line, and the assumption of hydrostatic equilibrium to derive the temperature as a function of column mass from observations. This method was used for late-type stars by \cite{1975ApJ...200..660A} and \cite{1974ApJ...192...93A}. It was also used for a sunspot penumbra by \cite{2002A&A...389.1020R}, who investigated its temperature structure using a scan obtained with the slit spectrograph at the former 50-cm Swedish Vacuum Solar Telescope, thus obtaining a 3D temperature model of the penumbra approaching 0\farcs23 resolution. That same data and method were again used by \cite{2005A&A...437.1069S} to compare the computed stratification obtained at different bright points, with predictions made for vertical two-dimensional thin fluxtube models. \cite{2012SoPh..280...83S} produced synthetic spectra from hydrodynamic (HD) and magneto-hydrodynamic (MHD) simulations and investigated the accuracy of a ``manual'' inversion technique for extracting temperature stratification, demonstrating good results for selected small structures.

Recent high-resolution work based on filtergrams recorded in the wings of these lines has focused on qualitative interpretation of images in the \ion{Ca}{ii}~H blue wing. \cite{2005A&A...431..687L} observed the inner blue wing of \ion{Ca}{ii}~H (396.667~nm) and compared it with line core filtergrams. \cite{2007ASPC..369...71S} and \cite{2007A&A...464..763L} also used an inner wing position to explore the upper photosphere of a sunspot penumbra. \cite{2004A&A...428..613B} observed in the G-band and close to the \ion{Ca}{ii}~H line core to characterize bright points and small pores at 0\farcs10  resolution. This last work used a 0.29 nm wide filter in a setup at the Swedish 1-meter Solar Telescope (SST) that is similar to the one used in the present paper regarding the secondary cameras.     

Few high-resolution semi-empirical 3D models of solar fine structure have been published. \cite{2011A&A...529A..37S}, recognizing the lack and the need for these in the context of abundance determination, used Hinode spectra in the 6301 and 6302 line region with the non-LTE (NLTE) inversion method of \cite{1998ApJ...507..470S} to produce a model of quiet granulation at disk center at 0\farcs3 resolution. 

At the SST, 0\farcs10 is achievable at the wavelength of the \ion{Ca}{ii}~H and K lines, and the use of filtergrams allows for control and compensation of seeing effects using image reconstruction techniques. In this work, we apply the method of \cite{1974SoPh...37..145S} to filtergrams in the blue wing of \ion{Ca}{ii}~H. We address the main image processing issues with such an adaptation of their method to filtergram data and test the method by producing and analyzing synthetic observations from an HD simulation. Finally, we compare observations and simulations and discuss different photospheric structures in different heights and in vertical temperature gradients.

\section{Setup and observations}
\label{sect:setup}

\begin{figure*}[t!]
  \centering
\resizebox{!}{13cm}{\includegraphics[clip=true]{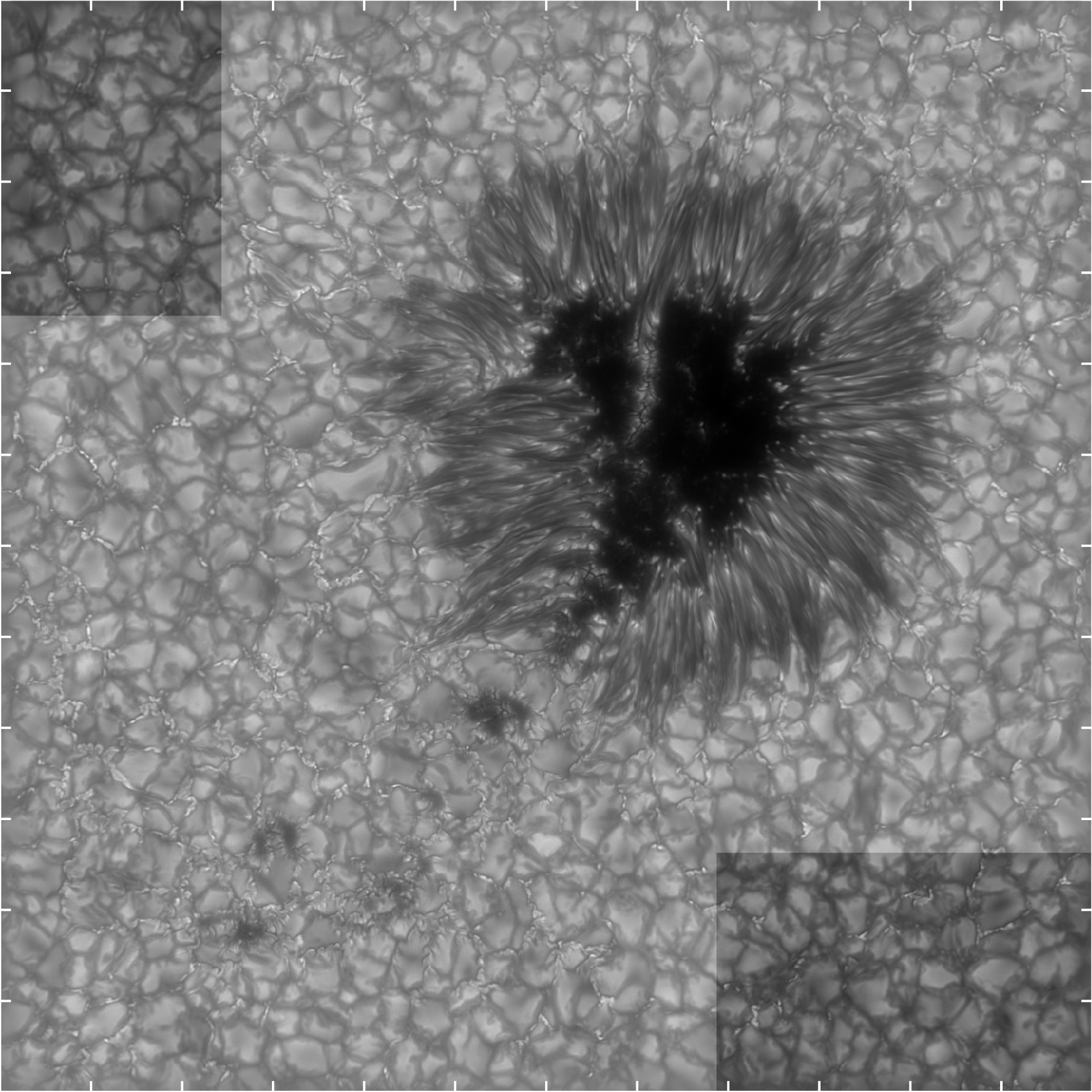}}
\caption{\footnotesize Wideband image of the whole field of view. Each tick mark is 5\arcsec. The highlighted ``quiet Sun'' areas in the upper left and lower right corners are used for intensity calibration.} 
\label{wideband}
\end{figure*}

The observations were performed on 23 May 2010 with the SST \citep{2003SPIE.4853..341S}, including the 37-electrode-mirror adaptive-optics system \citep{2003SPIE.4853..370S}. The target was active region NOAA 11072, including a sunspot, pores, and associated bright points (see Fig.~\ref{wideband}). The field of view is $62\arcsec \times 62\arcsec$ with an image scale of 0\farcs034 per pixel (25.0~km per pixel on the Sun). The active region was observed at a heliocentric angle of  $\theta\sim$15\degr corresponding to $\mu=\cos\theta=0.97$. 

The setup, known as the blue tower, included four Megaplus II es4020 cameras. The camera fed by the first beamsplitter cube, thus receiving the most light of the four cameras, was placed behind a 0.11 nm interference filter mounted on a rotation stage. The central wavelength at normal incidence of this filter is 396.88 nm, just red-wise of the core of \ion{Ca}{ii}~H, but rotating the filter shifts the center of the pass band towards the blue. This provides a simple way to scan the entire blue wing of the \ion{Ca}{ii}~H line at the cost of lower spectral resolution and image degradation for large tilt angles, as explained by \cite{1998A&AS..129..191B}. The second camera was mounted behind a 0.12 nm wide filter at 396.47 nm in the \ion{Ca}{ii}~H blue wing. The wing camera is included in the Multi-Object Multi-Frame Blind Deconvolution \citep[MOMFBD]{2005SoPh..228..191V} reconstructions but not in the present analysis because we focus on the tiltable filter. The last two cameras were mounted behind a wideband (WB) filter with a FWHM of 1 nm centered on 395.37 nm in the pseudo continuum between the \ion{Ca}{ii}~H and K lines. One of these was intentionally defocused by 7~mm for approximately one wave peak to peak of phase diversity (PD). 
  
The present data are from a single scan, which consists of 91 images per camera with the tunable filter camera scanning seven wavelengths in the \ion{Ca}{ii}~H blue wing plus line core (13 images per tunable filter wavelength). The scanned wavelengths and different filter-position properties are listed in Table \ref{table:1}. The exposure time was 8 ms, with the full scan taking 9.7 seconds.

The SST setup allows for co-observations in the red with the CRISP Imaging Spectropolarimeter \citep[CRISP]{2008ApJ...689L..69S}. The data, acquired at 14:11~UT, are nearly simultaneous with \ion{C}{i} 538.0~nm, \ion{Fe}{i}~630.2~nm, and 630.1~nm observations. Analyses of these observations are published in \cite{2011Sci...333..316S} and \cite{2012A&A...540A..19S}, where convective downflows were detected in the sunspots penumbra. We intend to combine the \ion{Ca}{ii}~H and CRISP data sets in future work.

\section{Data reduction}

Data reduction is centered around image reconstruction using the MOMFBD code \citep{2005SoPh..228..191V} with PD. MOMFBD was used with the 36 most significant Karhunen-Lo\`{e}ve coefficients. Since we aim at differentiating data, additional steps were taken 
to properly account for the complications that arise from the tiltable filter and to address frame non-simultaneity. The data reduction procedures are summarized below in chronological order and then discussed in detail in the specified sub-sections:

\begin{itemize}
 \item Non-linearity correction for one of the cameras (Sect.~\ref{sec:reduc1})
 \item Camera alignment with deconvolved pinholes (Sect.~\ref{sec:lowpass})
 \item Dark correction and flat fielding 
 \item MOMFBD with extra image objects (Sect.~\ref{dewarping})
 \item Compensation for tilted-filter PSF effects (Sect.~\ref{sec:lowpass})
 \item Dewarping using extra image objects (Sect.~\ref{dewarping})
 \item Noise filtering 	(Sect.~\ref{sec:lowpass})				
\end{itemize}

\subsection{Non-linearity correction and ghosts}
\label{sec:reduc1}

The MegaPlus II es4020 camera CCDs consist of two halves with separate read-out circuits. One of the halves of the PD camera showed a consistent non-linearity in the gain. After MOMFBD reconstruction, this difference between the two halves gave rise to a strong central artifact. This non-linear response was measured by Dr.~P.~S\"utterlin (2010 private communication) and was well modeled by a six-degree polynomial. The non-linear gain would introduce, at its peak, $\sim 6\%$ extra counts compared to the linear response curve from the other half. For this data set, that polynomial was used to correct each individual raw science and flat-field frame after dark correction for the affected camera. The camera was repaired between the 2010 and 2011 campaigns and no longer presents this issue.

In pinhole array images, we detected defocused ghost images of around 1\% for all positions of the tunable filter. Based on the progression of their distance to the main image with increasing tilt angle, these likely come from a secondary image generated by inner reflections in the tilted filter. We ignore these ghosts. They will have an impact similar to a small amount of scattered light. Extreme inverse filtering and inspection of the images in high contrast in the umbra do not reveal any visible ghosts in the final reconstructed images.

\begin{figure*}[]
  \centering
\resizebox{14.0cm}{!}{\includegraphics[bb=0 0 505 222,clip=true]{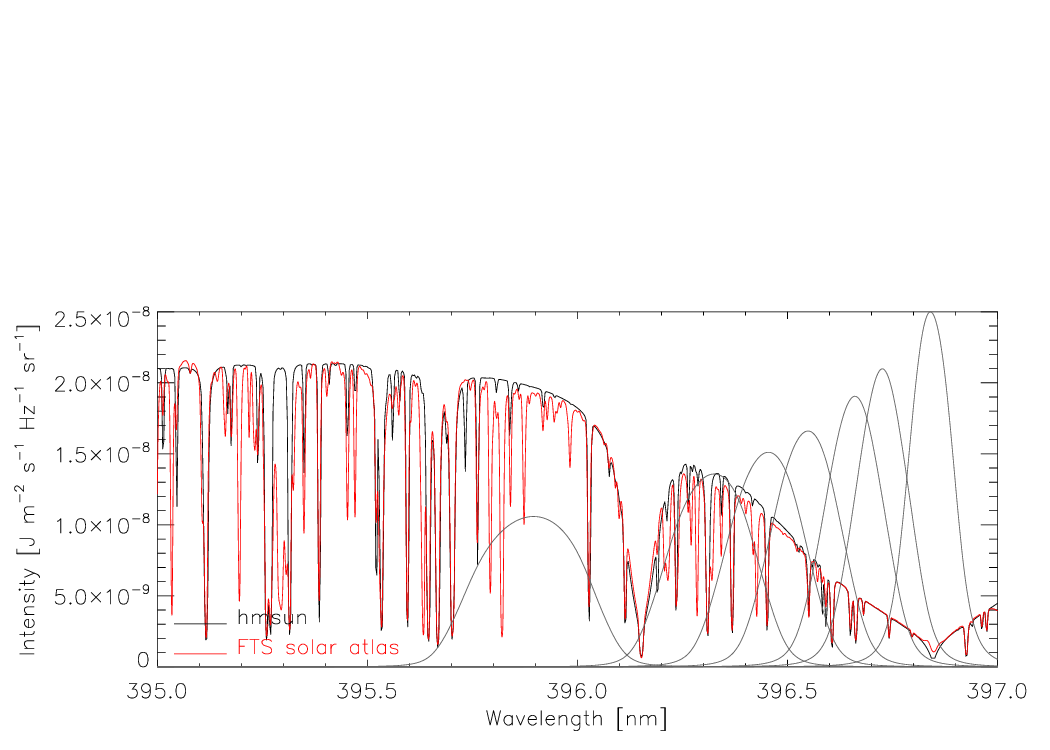}}

\caption{\footnotesize  Different transmission profiles ($F_i\left(\lambda\right)$) of the interference filter for the tilt angles used. The FTS atlas \citep{1999SoPh..184..421N} is shown in red, together with the synthetic spectrum from HolMul with blends (black continuous line) computed at $\mu = 0.953$. The filter profiles are normalized to the peak transmission of the narrowest profile in the plot, over the line core, that corresponds to the smallest tilt angle used. The large blend at 396.15 nm is from \ion{Al}{i}.}
\label{prof}
\end{figure*}

\subsection{Compensating for tilted-filter effects}
\label{sec:lowpass}

The degradation effects from tilting an interference filter in this setup are discussed in \cite{2011A&A...533A..82L}, where a compensation scheme in the context of MOMFBD reconstruction was also established. Here we apply that post-MOMFBD compensation scheme, which includes deconvolution of the reconstructed images with the known elongated point spread functions (PSFs) from tilting and extra noise filtering. This compensation is possible since the effects of a tilted filter are separable from phase aberrations and thus do not interfere with wavefront sensing in MOMFBD \citep{2011A&A...533A..82L}.  We follow the pre-MOMFBD pinhole deconvolution as proposed in that work and additionally apply a noise filter to the deconvolved pinhole images in the same way as for each tunable-filter science position prior to MOMFBD.

For the individual temperature maps, the images at each tilt-angle position were low-pass filtered with a different, individually optimized spatial filter, which was estimated from the noise levels of the final deconvolved images. However, when differentiating temperature maps (as done in Sect. ~\ref{sec:gradients}), the data to be included in the computation need to be filtered in the same way in order not to introduce artificial signal due to different resolutions of the two temperature maps. Since higher tilt angles require a more aggressive noise filtering (a lower pass threshold in the Fourier domain), we separate the data in two sets before differentiating data. For each set, the same noise filter is applied to all narrowband (NB) images. This filter is built by using the minimum, at each coordinate point in Fourier space, of the individually optimized noise filters. It is close to the noise filter from the highest tilt angle image used in each set. For one of the sets we use all the filter positions requiring aggressive filtering and limiting the achievable resolution to about 0\farcs17. For the other set, we use only the smallest four tilt angles so that the analysis of the gradients in the upper photosphere can be done at the maximum resolution (see Table~\ref{table:2} for the different sets). \\

\begin{table}
\caption{Filter properties at different positions.}             
\label{table:1}      
\centering          
\begin{tabular}{c c c c c c c}     
\hline\hline       
Pos. & Tilt  & $\lambda_{\mathrm{c}}$ & $FWHM$ & $\lambda_{\mathrm{m},i}$  & $m_i$ & $\lg\tau_{500}$    \\ 
 ($i$) & angle  & (nm) & (nm) & (nm) & ($\mathrm{g}\ \mathrm{cm}^{-2}$)  &   \\ 
\hline                    
   (1)&1\fdg2&396.84&0.12&396.81 & 0.17 & -2.62   \\  
   (2)&2\fdg5&396.74&0.14&396.71 & 0.90 &  -1.31  \\
   (3)&3\fdg0&396.67&0.16&396.64 & 1.40 &  -0.96  \\
   (4)&3\fdg7&396.57&0.18&396.51 & 2.18 &  -0.74 \\
   (5)&4\fdg2&396.47&0.19&396.43 & 2.61 &  -0.42 \\
   (6)&4\fdg8&396.34&0.22&396.33 & 3.04 &  -0.37 \\
   (7)&6\fdg4&395.93&0.28&395.88 & 3.56 & -0.27 \\
\hline                  
\end{tabular}
\tablefoot{The first column lists the position number as used in the text. $\lambda_{\rm{c}}$ is the central wavelength of the passband for each tilt angle and $\lambda_{\mathrm{m},i}$ the effective wavelength defined in Sect.~\ref{sub:intensity}. $\rm{m}_i$ is the average depth for the whole field of view computed using Eq.~\ref{eq5}, and $\lg\tau_{500}$ is the corresponding optical depth value in HolMul.}
\end{table}

\subsection{Making non-simultaneous frames consistent}
\label{dewarping}

\begin{figure}[!htb]

  \centering
\resizebox{!}{6cm}{\includegraphics[clip=true]{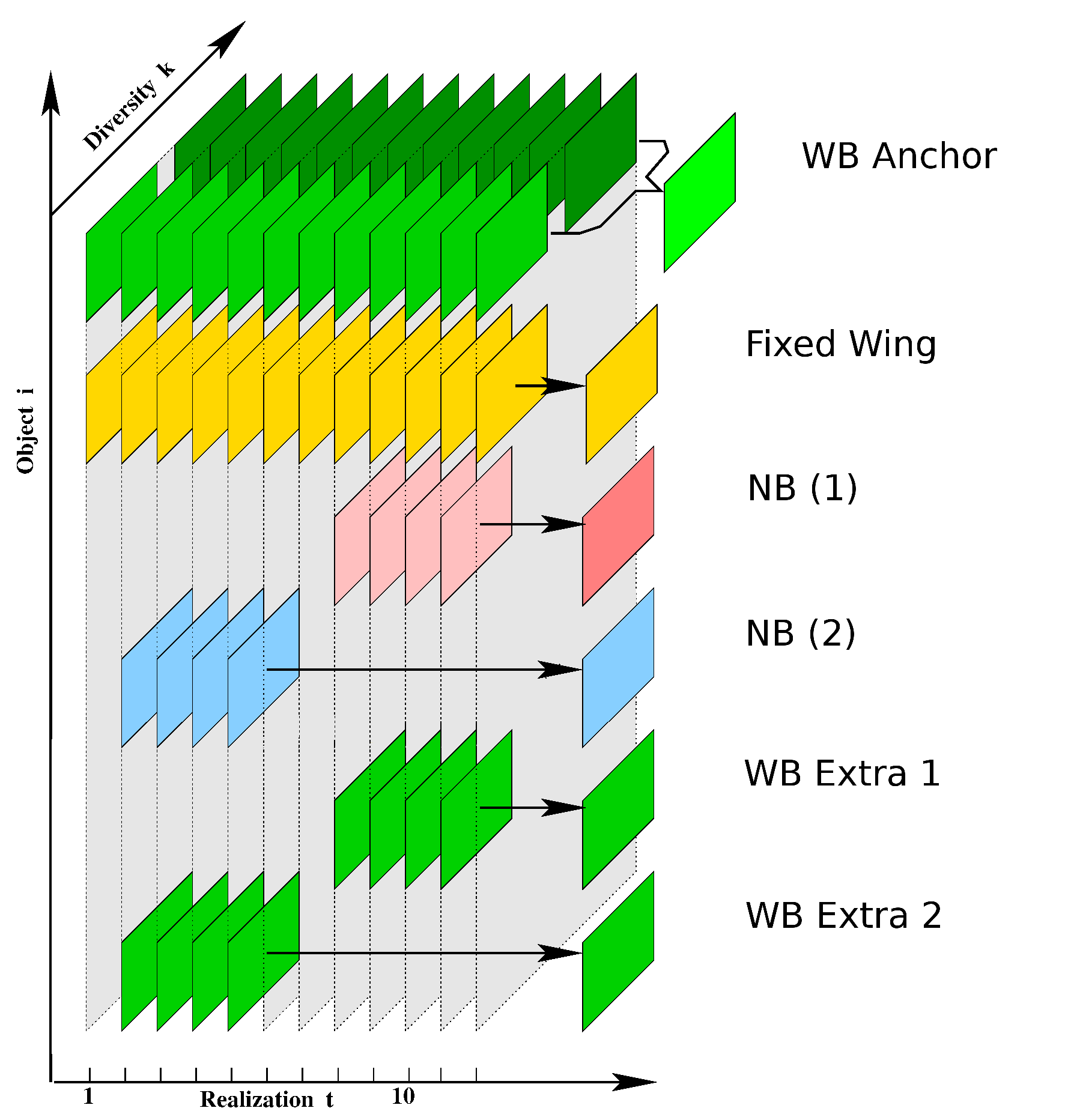}}
\caption{\footnotesize MOMFBD scheme with the extra WB objects, exemplified for two tunable filter positions. For each NB object (marked as ``NB''), a collection of simultaneous WB images (marked as ``WB Extra 1'' and ``WB Extra 2'') are selected and input as an extra object. The dark green images on top are from the WB PD camera. The scheme used for this work includes seven NB objects and their corresponding seven WB extra objects, each with 13 realizations.}

\label{scheme}
\end{figure}

It is not possible to perfectly compensate for seeing using image reconstruction. The different, non-simultaneous, tunable filter reconstructions will be affected by residual differential seeing effects (in particular, small-scale geometric distortions) that translate to false signal when differentiating data. However, it is possible to improve consistency between the non-simultaneous NB objects by using a simultaneous constant reference.  Such a reference is provided by the wideband object  (shown in Fig.~\ref{scheme} in light green). 

In order to take advantage of this reference, the MOMFBD reconstruction was done with one additional object (``WB'' in Fig.~\ref{scheme}) for each NB object consisting only of 
wideband frames that are simultaneous with individual frames for a particular NB object (``NB'' in Fig.~\ref{scheme}). The resulting reconstruction thus produces one WB image per NB reconstructed image that is affected by the same seeing and same reconstruction PSFs as the NB image. These individual WB images can then be used together with the reference (anchor) WB object spanning the full-scan time length to compute a geometrical distortion (``dewarp'') matrix obtained by using cross-correlation techniques to measure small-scale differential image motion.  Applying that dewarp matrix to the corresponding NB object improves the consistency between that NB object and all the other non-simultaneous NB objects, even though the procedure does not improve small-scale differential blurring. 

For this work the dewarping procedure involved applying three dewarp matrices in succession, with the last using 16 by 16 pixel boxes (0\farcs544) for correlation. This is approximately 5.5 times the diffraction limit (0\farcs10) that we believe is achieved in the WB objects used as well as for filter positions (1), (2), (3), and (4) for this data set. Outliers in each dewarp matrix were removed, the matrix was smoothed, and a box overlap of one-third was used. Due to the high contrast of the images and the abundance of visible small structure, perhaps even smaller boxes for correlation could be used. The extra WB objects were given zero weight in the MOMFBD wavefront determination. 

\begin{figure*}[!htb]
  \centering
\resizebox{!}{6cm}{\includegraphics[clip=true]{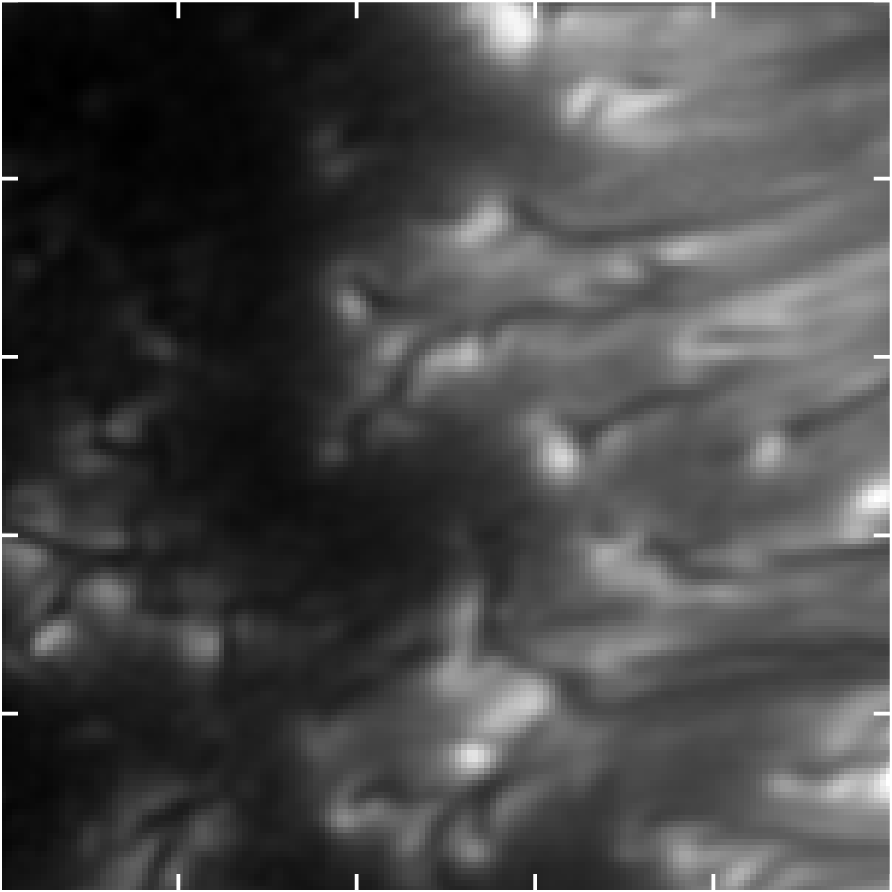}}
\resizebox{!}{6cm}{\includegraphics[clip=true]{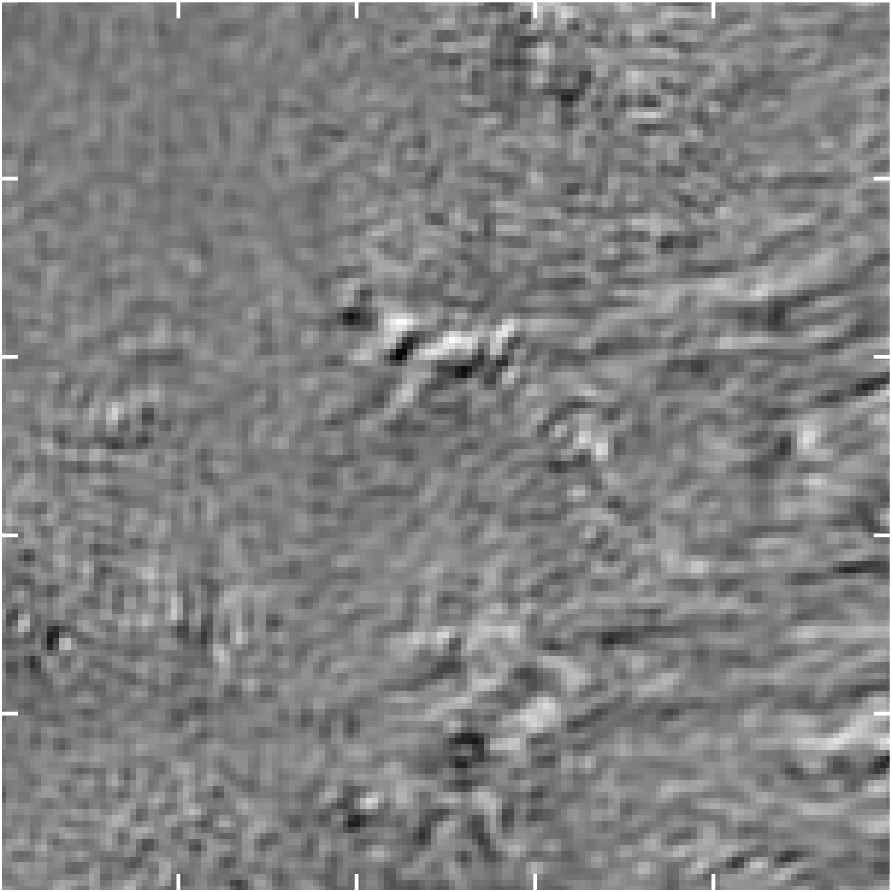}}
\resizebox{!}{6cm}{\includegraphics[clip=true]{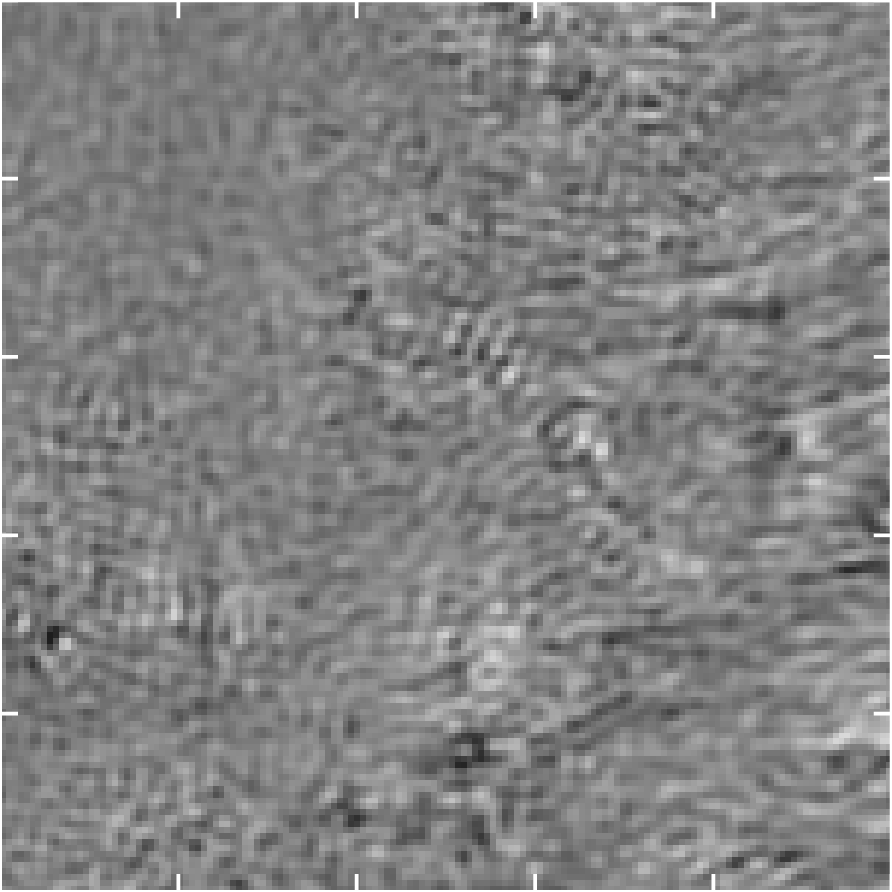}} 
\caption{\footnotesize Left: WB image. Center: difference between an unwarped WB extra object and the WB anchor object. Right: same as center but using the dewarped sub-object. Each tickmark is 1\arcsec. The scale in the center and right panels is the same.}
\label{dif}
\end{figure*}

To evaluate both the efficacy of this method, and provide a pixel-by-pixel measure of seeing-induced signal when differentiating data, we compute a difference map between each of the time-segmented WB objects and the anchor WB object. Using these difference maps, we can assess the amount of signal introduced by non-simultaneity and estimate how much false signal the dewarping procedure reduces. The full-field average improvement is small, as would be expected in good seeing conditions. However, there are marked improvements in areas that are related to high horizontal intensity gradients. This is the case for the filament heads shown in Fig.~\ref{dif}. Around the central filament, the full range of the difference map before the dewarping procedure is $-$4.5\% to 3.6\% of the WB dynamic range and improves to $-$3.6\% to 2.5\% after dewarping. 

When computing temperature gradient maps, we used data compensated in this way. When looking at maps from single images, we apply no dewarping (such as in the intensity maps of Fig.~\ref{intonly} and Fig.~\ref{tmaps}). 

\section{Calibration and inversion method}

\subsection{Intensity calibration}
\label{sub:intensity}

The counts of each pixel were converted to absolute intensity using a synthetic profile computed in NLTE using the line-synthesis code MULTI \citep[version 2.3]{MatsUppsala1986}. The Holweger-M\"{u}ller \citep[HolMul]{1974SoPh...39...19H} model atmosphere is used because this well-tested empirical model is constructed to reproduce spectral
lines in LTE. Blends were included in LTE from the VALD database \citep{1995A&AS..112..525P,1997BaltA...6..244R,2000BaltA...9..590K,1999A&AS..138..119K}. We also include the \ion{Ca}{ii}~K line as a background opacity for the full wavelength range. The obtained profile matches the National Solar Observatory Fourier Transform Spectrometer (FTS) atlas \citep{1999SoPh..184..421N} reasonably well (see Fig.~\ref{prof})  and provides information about the opacity contribution from \ion{Ca}{ii}~H and the background sources at different wavelengths. Profiles were computed for a range of $\mu$ values, including that of the observations. For the observational quiet Sun reference, we selected two regions at opposite extremes of the field of view as indicated in Fig.~\ref{wideband}. These regions were chosen for their relatively low activity using an image from an inner wing filter position. An intensity calibration factor per filter position was then computed from these regions and the calibration synthetic spectra. For clarity, the expression for the calibrated observed intensity is

\begin{equation}
		I_{i} = \frac{N_i}{\bar{N}_{i,\mathrm{QS}}}  \frac {\int I_i^\mathrm{s}(\lambda) F_i\left(\lambda\right)  \mathrm{d}\lambda }{\int F_i\left(\lambda\right)   \mathrm{d} \lambda} \,, 
 \label{inten}
\end{equation}
where $i$ is the filter position, $N_i$ is the number of counts in the observed pixel, $\bar{N}_{i,\mathrm{QS}}$ is the mean of the selected quiet Sun pixels, $F_i\left(\lambda\right)$ is the filter transmission profile (see Fig.~\ref{prof}) computed from the tilt angle as in \cite{2011A&A...533A..82L}, and $I_i^\mathrm{s}\left(\lambda \right)$ is the synthetic spectrum.

\subsection{Temperature and depth estimates}
\label{sub:depth}

Each filter position only provides one temperature and one wavelength point per pixel. The asymmetric intensity distribution of the line over each filter profile will cause the observed data to have effective central wavelengths that are blue-ward of the central wavelength values of each passband. For this reason, we multiply the filter profiles with the synthetic spectrum and use the wavelength of maximum intensity ($\lambda_{\mathrm{m},i}$) when determining temperature, depth, and background opacities. This is verified to be at a wavelength not dominated by a blend. 

We compute the brightness temperature ($T_b$) using the calibrated intensity ($I_i$) and $\lambda_{\mathrm{m},i}$:  

\begin{equation}
	  T_\mathrm{b} = B^{-1}\left(I_i,\lambda_{\mathrm{m},i}\right)
	  \label{trad} 
\end{equation}
where $B^{-1}$ is the inverse Planck function. Then, assuming $T_{kin}=T_b$, we assign a column mass density for each pixel using expressions (4) and (5) from \cite{2002A&A...389.1020R}, who clarified the expression of \citet{1974SoPh...37..145S}, and simply adapted to \ion{Ca}{ii}~H:

\begin{equation}
\begin{split}
		N_\mathrm{H}\left(\Delta\lambda\right) = &  \frac{1}{2} \left( \frac{5000}{T_\mathrm{b}} \right)^{0.39}  \left[ -\frac{\gamma_{\mathrm{rad}}}{\Gamma_\mathrm{W}} + \Bigg{\{} \left( \frac{\gamma_{\mathrm{rad}}}{\Gamma_\mathrm{W}}  \right)^2 + \right.  \\ 
		           & \left.   \frac{4g\mu\mu_{\mathrm{mol}}m_\mathrm{H}}{A_{\mathrm{Ca}}\Gamma_\mathrm{W} \textit{k}} \frac{m_\mathrm{e} 4 \pi c^3 q_{\lambda}}{e^2 f_{\mathrm{lu}}\lambda^4_0}\left( \frac{T_\mathrm{b}}{5000} \right)^{0.39} \frac{\Delta\lambda^2}{T_\mathrm{b}} \left( 1 - \tau_{\mathrm{c}}\right) \Bigg{\}}^{1/2} \right]   
\end{split}
\label{eq4} 
\end{equation}
 
\begin{equation}
		m = 1.1 N_\mathrm{H} \textit{k} T_\mathrm{b} g^{-1} 
		\label{eq5}
\end{equation}
with an oscillator strength of $f_{lu}=0.3155$ and radiative damping constant of $\gamma_\mathrm{rad} = 1.42  \times 10^{8}\ \rm{s}^{-1}$, both from \cite{1989PhRvA..39.4880T}. The coefficient $\Gamma_\mathrm{W} = 1.33 \times 10^{-8}$ is calculated by \cite{2002A&A...389.1020R} from the broadening cross-section of \cite{1998MNRAS.300..863B}. The neutral gas mean molecular weight is taken to be $\mu_{\mathrm{mol}}=1.29$. Abundances, including that of Ca~($A_{\mathrm{Ca}}$ above) and for the elements used in MULTI, are taken from \cite{2009ARA&A..47..481A}. The continuum optical depth correction $\left(1-\tau_{\mathrm{c}}\right)$ was computed with MULTI using the background optical depth ratio to total opacity at $\tau(\lambda_{\rm{m},i})=\mu$ in the HolMul model. The core wavelength is taken to be $\lambda_0=396.846$ nm and $\Delta\lambda=\lambda_0-\lambda_{\mathrm{m},i}$. We use the one-point quadrature weight $q_{\lambda}=1.6$ as defined and estimated by \cite{2002A&A...389.1020R}. Finally, $m_H,g,e,m_e,k,c$ are the hydrogen's mass, solar surface gravity, electron charge, electron mass, Boltzmann's constant, and speed of light, respectively. These equations assume LTE, and hydrostatic equilibrium and do not account for magnetic forces. 

The temperature dependence of the depth scale leads to each pixel being assigned a slightly different column mass. In order to compute temperature maps for a uniform depth-per-filter position, the temperature stratification per pixel, as given by the seven images and the equations above, is interpolated with a cubic spline to a common depth scale. 

\subsection{Temperature gradient maps}
\label{sub:grad}

Each gradient map in Fig.~\ref{intonly} and Fig.~\ref{tmaps} was computed from two temperature maps:

\begin{equation}
		G_{ij} \left(x,y\right) = \frac{T_{m_i}\left(x,y\right)-T_{m_j}\left(x,y\right)}{m_i - m_j} 
		\label{eq6}
\end{equation}

However, since the temperature map for each depth was interpolated to a single mean depth using a bicubic spline including the four nearest neighbors, gradients do include some information from other sampled wavelengths and smooth out the quantities by a small amount. In Table~\ref{table:2} we show the depths corresponding to the displayed temperature gradient maps and the temperature map pairs used to compute them. The images used to compute the temperature maps are filtered using two different sets of noise filters as described in Sect.~\ref{sec:lowpass} and identified in Table~\ref{table:2}. 

\begin{table}
\caption{Shown gradient maps, depths}             
\label{table:2}      
\centering          
\begin{tabular}{c c c c c c }     
\hline\hline       
                    
Number & Using $T$ maps  & $m$ & Resolution &  \\ 
 &  from positions  & ($\mathrm{g\ cm^{-2}}$)  &  & \\ 
\hline          
   $\mathrm{G_{12}}$ & (1+),(2+)   & 0.54  & 0\farcs10   \\    
   $\mathrm{G_{23}}$ & (2+),(3+)   & 1.15  & 0\farcs10   \\
   $\mathrm{G_{34}}$ & (3+),(4+)   & 1.78  &  0\farcs10   \\  
   $\mathrm{G_{45}}$ & (4-),(5-)  & 2.39  &  0\farcs17   \\ 
   $\mathrm{G_{56}}$ & (5-),(6-)  & 2.82  &  0\farcs17   \\
   $\mathrm{G_{67}}$ & (6-),(7-)   & 3.30 &  0\farcs17   \\
   $\mathrm{G_{57}}$ & (5-),(7-)   & 3.09  &  0\farcs17   \\
\hline                  
\end{tabular}
\tablefoot{The first column is an identifier for the gradient map as used in the text and figures. The second column identifies the filter positions used. The ``-'' sign identifies images filtered with the low-resolution noise filter and the ``+'' sign images filtered with the high-resolution noise filter. The third column ($m$) is the depth in column mass. The last column is the maximum resolution allowed by the noise filter used.}
\end{table}

\subsection{Synthetic data}
\label{syntss}
Synthetic data were produced from two different simulations. To test the method and assist interpretation of  non-magnetic areas in the observations, we use a 3D HD simulation snapshot computed with the code of \cite{1998ApJ...499..914S} and kindly provided by Dr.~M.~Carlsson. To assist interpretation of areas containing numerous bright points and pores, we use a 3D MHD simulation snapshot from Lagerfj\"{a}rd \& Nordlund (in prep.), kindly provided by Dr.~\r{A}.~Nordlund. This simulation has a mean total magnetic field of approximately 560 G around $\lg\tau_{500}=0$ for the section shown in Fig.~\ref{tmaps}. 

In each case, we again use MULTI to compute a synthetic profile with one plane-parallel atmosphere per pixel  (``1.5D geometry''), using a six-level \ion{Ca}{ii} model atom and the same parameters (such as abundances, oscillator strength, and damping constant) used for the HolMul reference spectrum and Eq.~(\ref{eq4}) and including the same line blends. We take the same $\mu$ as the observations. Synthetic observations are then produced by multiplying the synthetic line profiles with the different filter profiles. The total area of the simulations is 6x6~Mm for the HD simulation and 12x12~Mm for the MHD simulation. Both snapshots extend up to about 500~km above the mean continuum-forming layer of $\left<\lg\tau_{500}\right>=1$ with the HD simulation extending to 2.5~Mm below and the MHD simulation 20~Mm below the surface. Neither simulation has a chromosphere.  The number of pixels from the MHD simulation is reduced by a factor of four such that the pixel scale is 23.8~km per pixel for both simulated snapshots. This is very close to the pixel size of the observations (see Sect.~\ref{sect:setup}) so we do not interpolate the synthetic data to a new scale. 

When comparing observations and simulations (see Sect.~\ref{sec:gradients}), it is necessary to degrade the synthetic images to the quality of the observations. It is difficult to determine an exact degradation PSF to apply to the synthetic images. The diffraction by the finite telescope aperture as well as a low-order representation of the phase aberrations are already included in the deconvolution performed in MOMFBD, and we deconvolve our observational data using the different theoretical interference filter PSFs as described in Sect.~\ref{sec:lowpass}. Furthermore, some aspects of the image degradation (in particular straylight) are poorly known. However, the final reconstructed observations are noise filtered in a relatively aggressive way, effectively reducing the maximum achievable resolution from 0\farcs1 to 0\farcs17 for the largest tilt angles (as described in Sect.~\ref{sec:lowpass}). The fact that the noise filters are based solely on the power spectrum of the actual final reconstructed images means that they include unknown image-degradation effects that are not compensated for in the observations. Therefore, for the intensity maps in Sect.~\ref{sec:gradients}, we use the tilt-angle-specific noise filters and, when calculating temperature gradients, we use two different sets of noise filters (see Sect.~\ref{sec:lowpass}), as done in the observations.

\begin{figure}[] 
  \centering
\resizebox{!}{8cm}{\includegraphics[trim=1cm 4cm 1cm 0,clip=true, totalheight=0.8\textwidth]{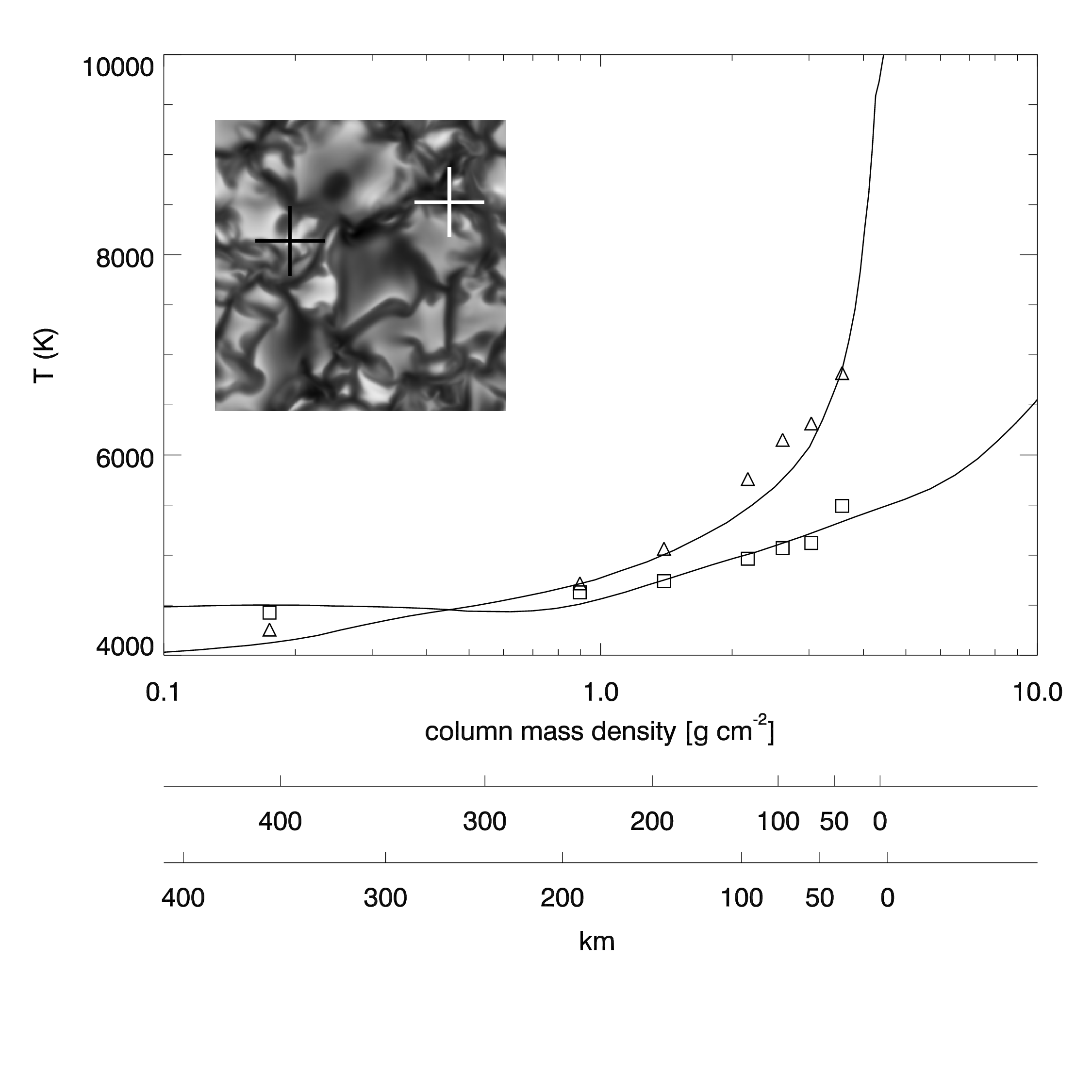}}
\caption{\footnotesize Temperature stratification for two columns, extracted from the synthetic images, are shown. The solid lines are the respective known temperature stratifications from the 3D HD simulated atmosphere. The dark crosshair marks the brightest near-continuum pixel and corresponds to the upper set of points, upper line, and the upper geometrical height scale. The white crosshair marks the darkest pixel and corresponds to the bottom set of points, bottom line, and bottom geometrical height scale, with zero defined by the mean optical depth $\left<\lg\tau_{500}\right>=1$.}
\label{the2dplot}
\end{figure}

In addition to the above filtering, we add straylight to the synthetic images in the same way and with the same parameters as \cite{2011Sci...333..316S}, using a 1\farcs2-wide Gaussian straylight PSF and considering a 58\% straylight component at a wavelength of 538~nm. The large difference in wavelengths between these two sets of observing data, in addition to the uncertainties already associated with this choice of straylight parameters, means that this is will not be the correct straylight function. However, our goal is only to make a morphological comparison between observations and synthetic data, and we do not attempt to remove straylight from the observations. The maps shown in Fig.~\ref{intonly} and Fig.~\ref{tmaps} are computed from synthetic images degraded in this way (including straylight).  

The root mean square contrast (RMS) of the observed position (7) quiet Sun is approximately that of the WB object (which is 12.5\%). After introduction of straylight, the RMS contrast of the synthetic granulation images is still slightly higher than that of the observations (see Table~\ref{table:RMS}). The value of the degraded near-continuum position (7) is consistent with comparable values from both synthetic and observed data discussed by \cite{2009A&A...503..225W}.

\begin{table}
\caption{\footnotesize RMS of the observational intensity versus the HD synthetics.}             
\label{table:RMS}      
\centering          
\begin{tabular}{c c c c c}    
\hline\hline       
Pos. & Obs.  & Obs.  & HD & HD   \\ 
 ($i$) &  MOMFBD &  filtered &   & degraded  \\ 
\hline                    
   (1) & 27\% & 18\% & 38\% & 19\%  \\
   (2) &  18\%  & 13\% & 34\% & 16\%  \\ 
   (3) &  15\% & 11\% & 27\% & 13\%  \\
   (4) &  12\% & 10\% & 20\% & 11\%  \\
   (5) & 11\%  &  10\% & 22\% & 12\%  \\
   (6) &  11\% & 10\% &  25\% & 13\%  \\
   (7) &  12\%  & 12\% &  30\% & 14\%  \\
\hline                  
\end{tabular}
\tablefoot{Columns in order: filter position, RMS of the quiet Sun patches shown in Fig.~\ref{wideband}, RMS of the same patches after the most aggressive filtering (Sect.~\ref{sec:lowpass}), RMS of the synthesized HD images, and RMS of the degraded HD synthetics, including straylight as described in Sect.~\ref{syntss}.}
\end{table}

We assume, for our aims, that the combination of this noise filtering and straylight model includes a reasonably realistic representation of the unknown image-degrading effects present in the observations. The fact that maps from both synthetic and observational data (in Fig.~\ref{intonly} and Fig.\ref{tmaps}) have the same dynamic range is a good indication of this and allows us to compare the morphology in the different computed quantities (see Sect.\ref{sec:gradients}).

\section{Testing with a known 3D atmosphere} 
\label{section5}

To test our procedure, we use the undegraded synthetic images produced from the field-free HD simulation as described in Sect.~\ref{syntss}. We extract the temperature from the synthetic data using the methods in Sect.~\ref{sub:intensity}~and~\ref{sub:depth} and plot it against the actual temperature from the input simulation snapshot, at the same column mass for each filter position in Fig.~\ref{hetero2}. Mean error and number of pixels below a certain error threshold are listed in Table \ref{table:errors}. To illustrate the procedure in detail, a plot of temperature as a function of height with all the filter positions is shown for the coldest and hottest pixels in Fig.~\ref{the2dplot}.  
 
There are no tilt angles or temperature ranges where the method fails entirely. The extracted temperature generally follows the actual temperature from the simulation atmosphere. 

The errors are highest for the largest filter tilt angle (7) with an average over 300 K (see Table~\ref{table:errors}). The largest errors for this tilt angle correspond to pixels with very high vertical temperature gradient, where small errors in the depth determination will cause a big mismatch in temperature. Furthermore, this is the deepest position and thus has the highest overall temperature gradient. Uncertainties in the derived depth at this tilt angle include the filter transmission profile being the broadest, \ion{Ca}{ii} no longer being predominantly in the ground state (with the level population varying rapidly with temperature), and the continuum opacity being of significant importance. These last two aspects are part of the limitation of the several one-point (height independent) assumptions in this simple analytical approach. Further one-point assumptions include the constant mean molecular weight and the one-point quadrature weight for the pressure integral ($q_{\lambda}$). We note that the continuum-opacity correction is computed for every filter position, but not as a function of temperature. The filter broadness introduces errors by mixing information from different layers and by shifting the effectively sampled wavelength blue-ward of the fixed $\lambda_\mathrm{m,i}$ in high-gradient areas and red-ward in low-gradient areas. The blue-ward shifting effect is expected to lead to an offset of temperature in the same direction of the offset observed in Fig.~\ref{hetero2}.

The width of the filter also leads to a much narrower sampled height range than for spectroscopic data by causing the smallest filter tilt position (1) to have a greater contribution from the line wings than from the line core. The width also limits the number of relevant sampling points. In Fig.~\ref{intonly}, chromospheric information is visibly mixed with the upper photosphere at position (1). 

The second worst position is position (5) with a mean error of 141 K. This position shows the same correlation of errors with high vertical temperature gradients as discussed previously, so the dominant error sources are probably the same as for position (7). The reason why position (6) is not worse than position (5) is that it is heavily contaminated with the \ion{Al}{i} line (as visible in Fig.~{\ref{prof}), causing its formation height to be higher than what is calculated using Eq.\ref{eq5}. This partially compensates the offset error from the depth-independent quantities.

For all filter positions except (5) and (7), the mean error is below 100 K and more than half the pixels are below 100 K error. Somewhat surprisingly, for every filter position except position (5), the errors are smallest for the extremes of the temperature range. 

The MHD simulation is much colder than HolMul in the middle-to-upper photosphere. Due to the requirement of intensity calibration with the assumption that the observed quiet Sun regions will have the same average profile as the one computed from HolMul, testing this method with the MHD synthetic data leads to large offset errors and a generally large error scatter that may or may not be related to the intensity calibration. For this reason, we decided to focus on the HD test. Nonetheless, it is relevant to mention that generally, just like in the HD case, the error for the temperature extremes for each filter position is relatively small. This means that the temperature-extraction method seems to work well for pores and bright points. 

A similar test to the one shown here using simulation data and the \ion{Ca}{ii}~K and H lines is performed by \cite{2012SoPh..280...83S}. She tested a manual fitting technique in full-resolution synthetic spectra and performed a test in selected columns from simulated 3D atmospheres and another test in 1D atmospheres averaged from those atmospheres. In the latter, she encountered problems mainly due to the non-linear temperature response of the Planck function, an issue that will always be present in unresolved structures but is not very important at our high spatial resolution. In the resolved test, she encountered problems where the temperature gradient is sharp, similar to what we see. The fact that it is a manual technique precludes its use with filtergrams and, as the author points out, an automatic technique is desirable (also due to the sensitivity of a manual method to the initial assumed atmosphere).

\begin{figure*}[!htb]
  \centering
\resizebox{!}{5cm}{\includegraphics[clip=true]{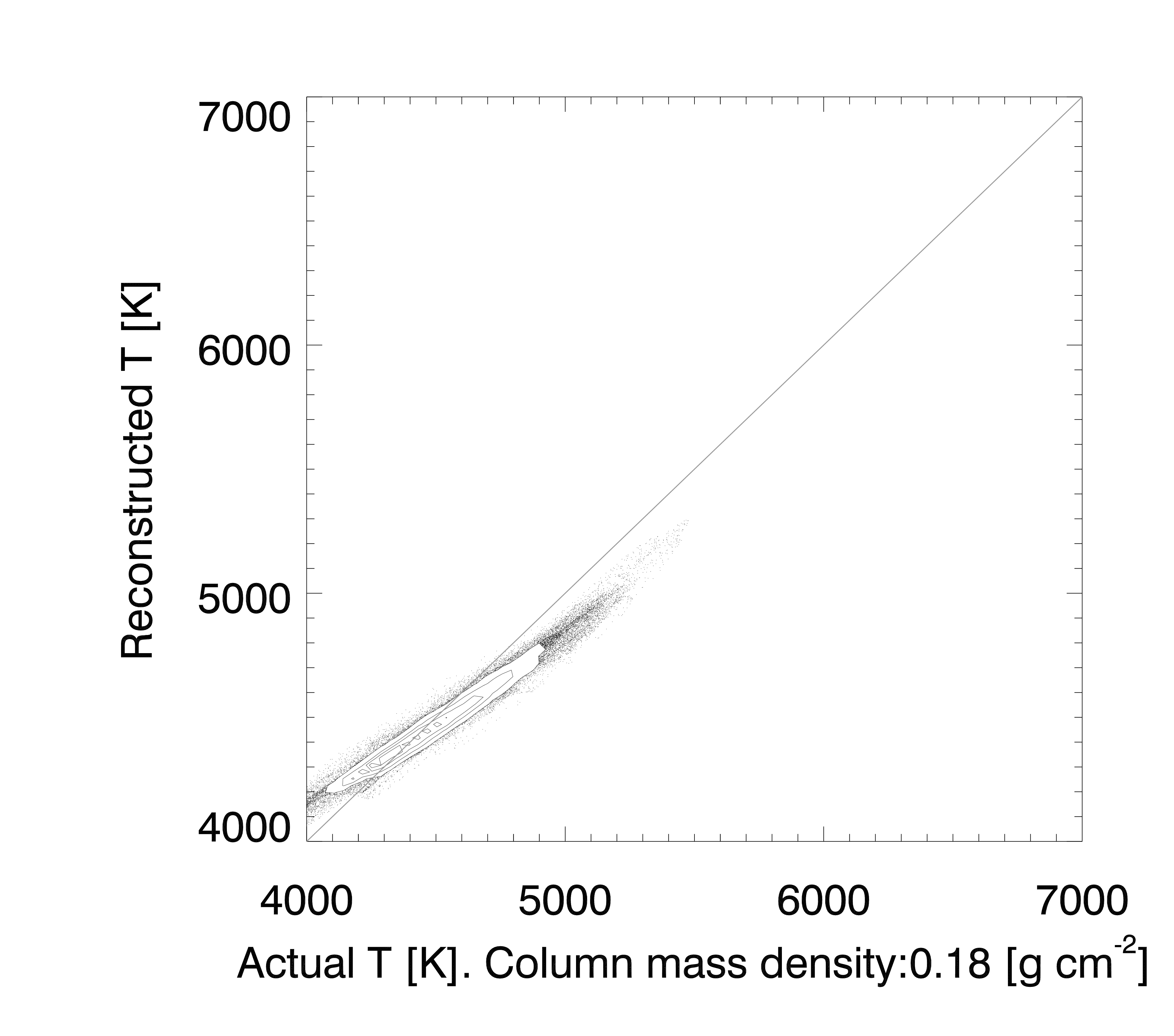}}
\resizebox{!}{5cm}{\includegraphics[clip=true]{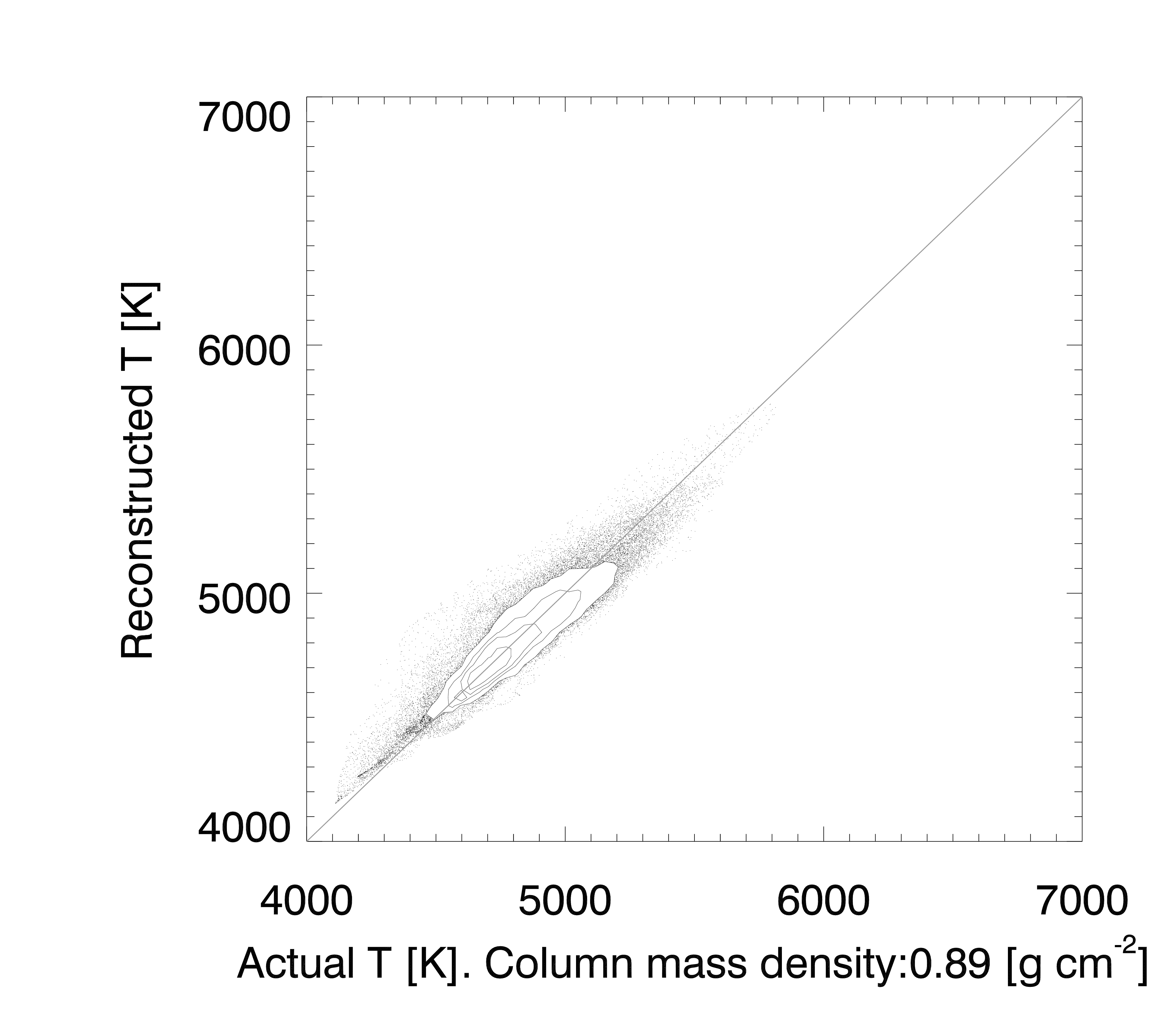}}
\resizebox{!}{5cm}{\includegraphics[clip=true]{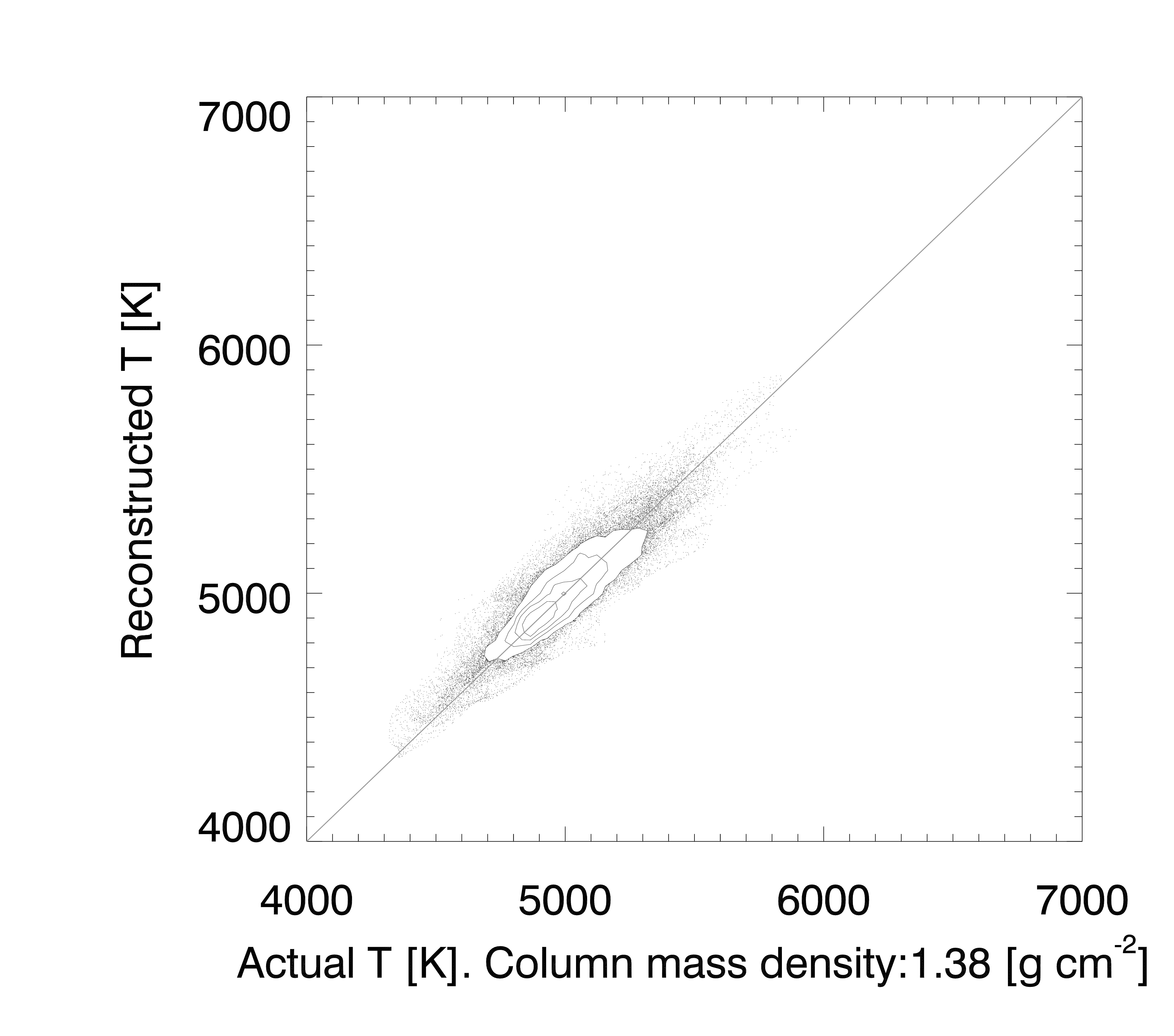}}
\resizebox{!}{5cm}{\includegraphics[clip=true]{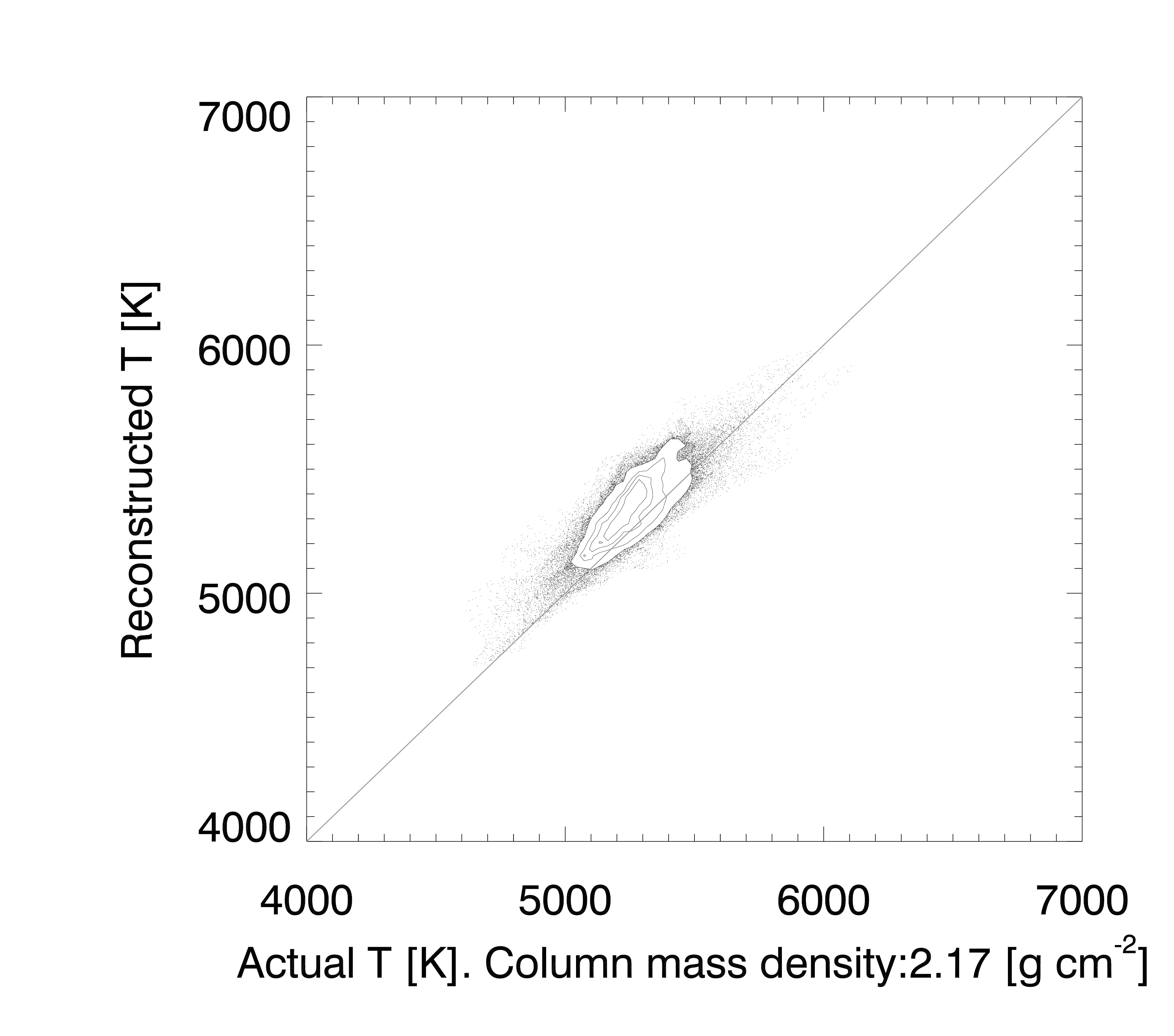}}
\resizebox{!}{5cm}{\includegraphics[clip=true]{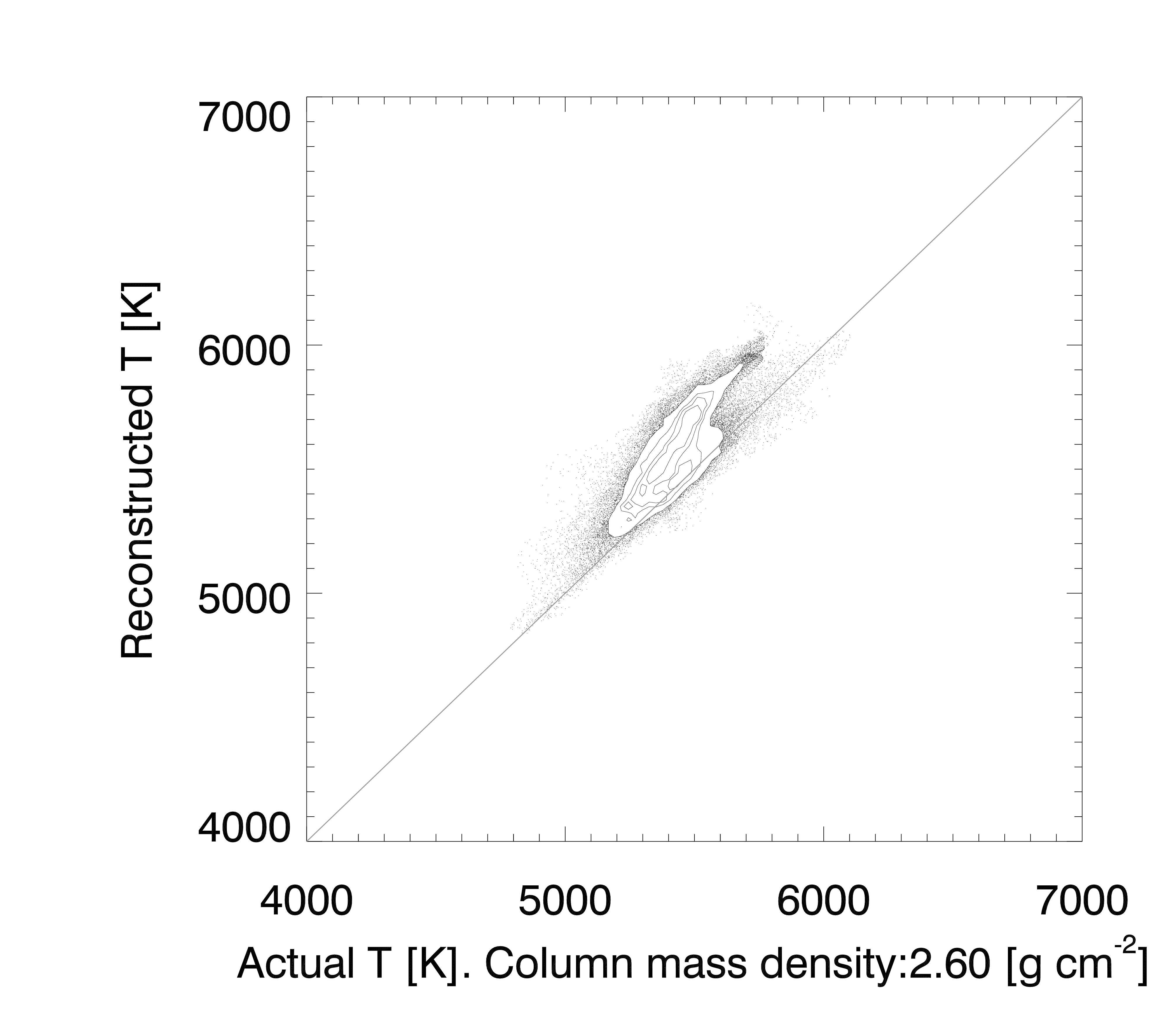}}
\resizebox{!}{5cm}{\includegraphics[clip=true]{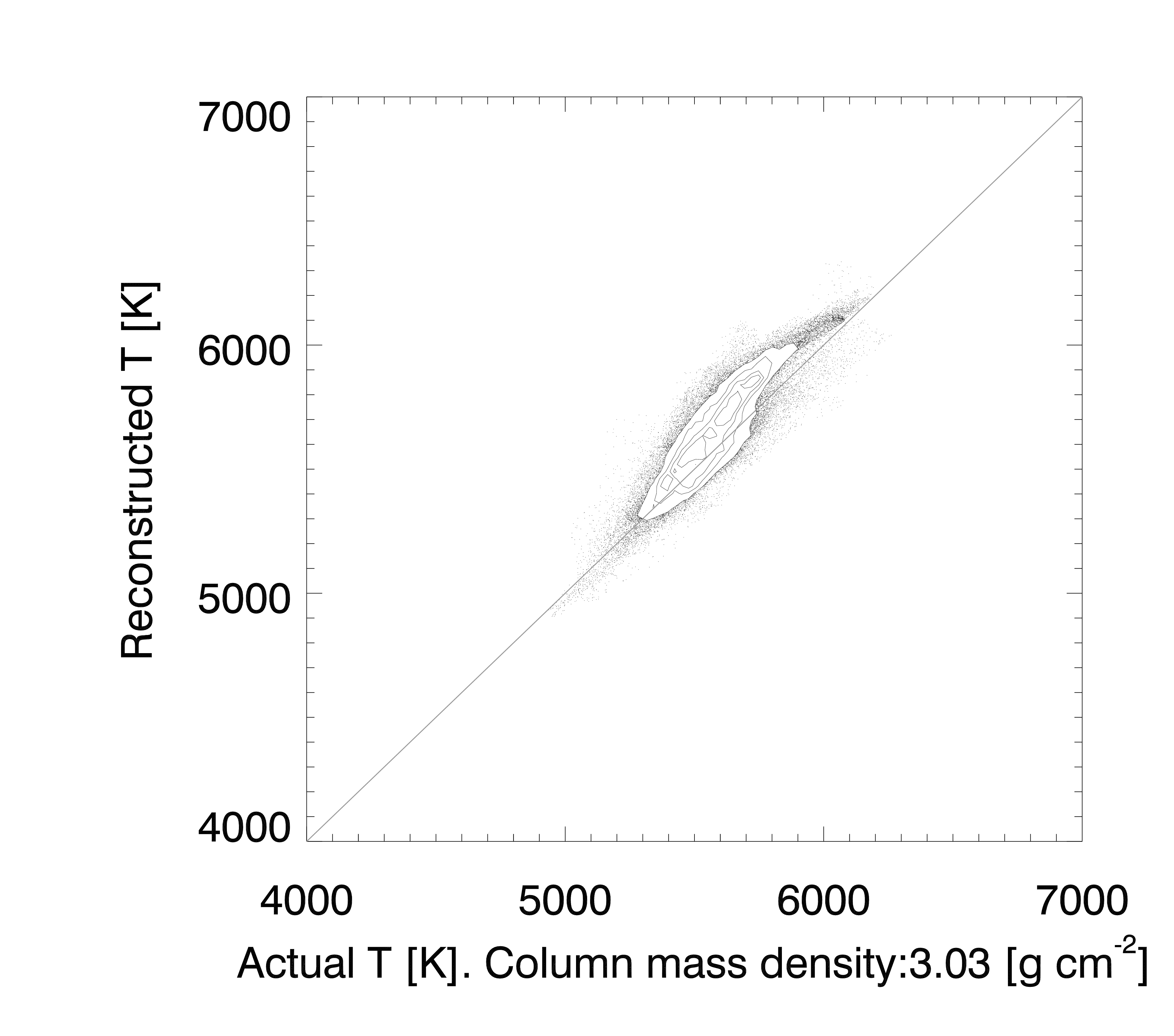}}
\resizebox{!}{5cm}{\includegraphics[clip=true]{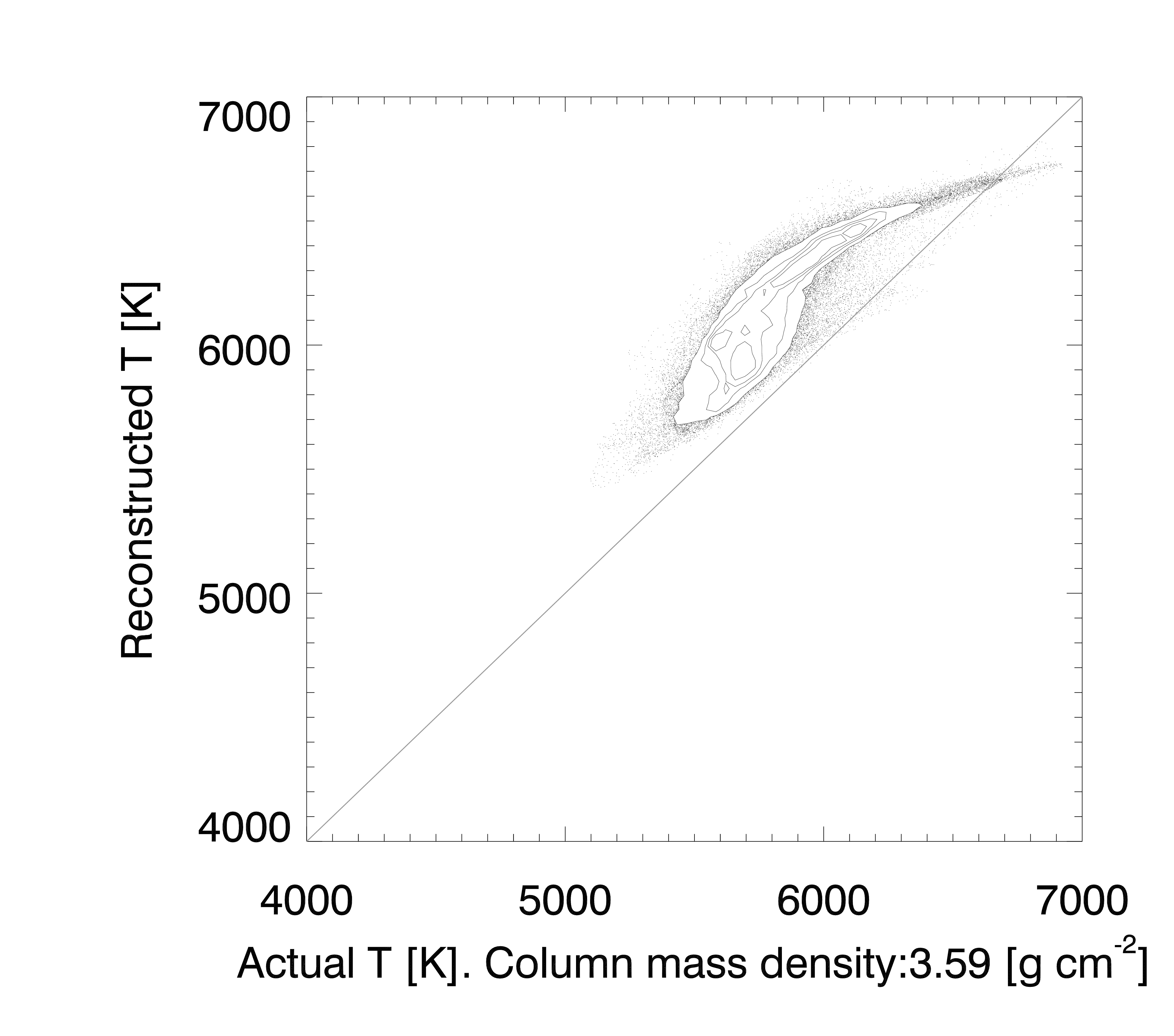}}

\caption{\footnotesize Known simulated temperature from the 3D HD snapshot versus temperature extracted from the synthetic observations. The contour levels, from the center of the clouds to the periphery, correspond to 20\%, 40\%, 60\%, and 80\% of the number of points.} 
\label{hetero2}
\end{figure*}

\begin{table}
\caption{Results of the test in Sect.~\ref{section5}.}             
\label{table:errors}      
\centering          
\begin{tabular}{c c c c c}     
\hline\hline       
Pos. & $<$error$>$  & error $< 50 K$ &  error $< 100 K$  & error $< 100 K$ \\ 
 ($i$) & (K) &  &  & minus offset   \\ 
\hline                    
   
      (1)& 73 & 0.42  & 0.73 & 0.75 \\  
   (2)&  61 & 0.51  & 0.81 & 0.81  \\
   (3)& 60 & 0.52 & 0.82 & 0.82 \\
   (4)& 94 &  0.30  & 0.58  & 0.76 \\
   (5)& 141 &  0.22 & 0.38   & 0.62  \\
   (6)& 94 &  0.29 & 0.59  & 0.71 \\
   (7)& 323 & 0.02  & 0.04  & 0.54  \\
\hline                  
\end{tabular}
\tablefoot{The first column lists the position number as used in the text. The second column is the mean error from the test described in Sect.~\ref{section5} and plotted in Fig.~\ref{hetero2}. The third and fourth columns are the fraction of pixels below 50K and 100K of absolute error, respectively. The fifth column is the same as the fourth, but after a fixed temperature offset has been subtracted.}
\end{table}

\begin{figure*}
  \centering
  \def\lab#1{\begin{minipage}[b]{14mm}
      #1\vspace{15mm}
    \end{minipage}}
  \def\tile#1{\resizebox{!}{3.3cm}{ \includegraphics{#1}}}
\begin{tabular}{c@{\hspace{2mm}}c@{\hspace{1mm}}c@{\hspace{2mm}}c}

       &  {\hspace{2mm}}Synthetic Intensity {\hspace{6mm}} Observational Intensity & {\hspace{2mm}}Synthetic Gradient {\hspace{6mm}} Observational Gradient   &   \\
       \lab{\centerline{(1)} 0.17~$\mathrm{g\ cm^{-2}}$}  &   \resizebox{!}{3.3cm}{ \includegraphics{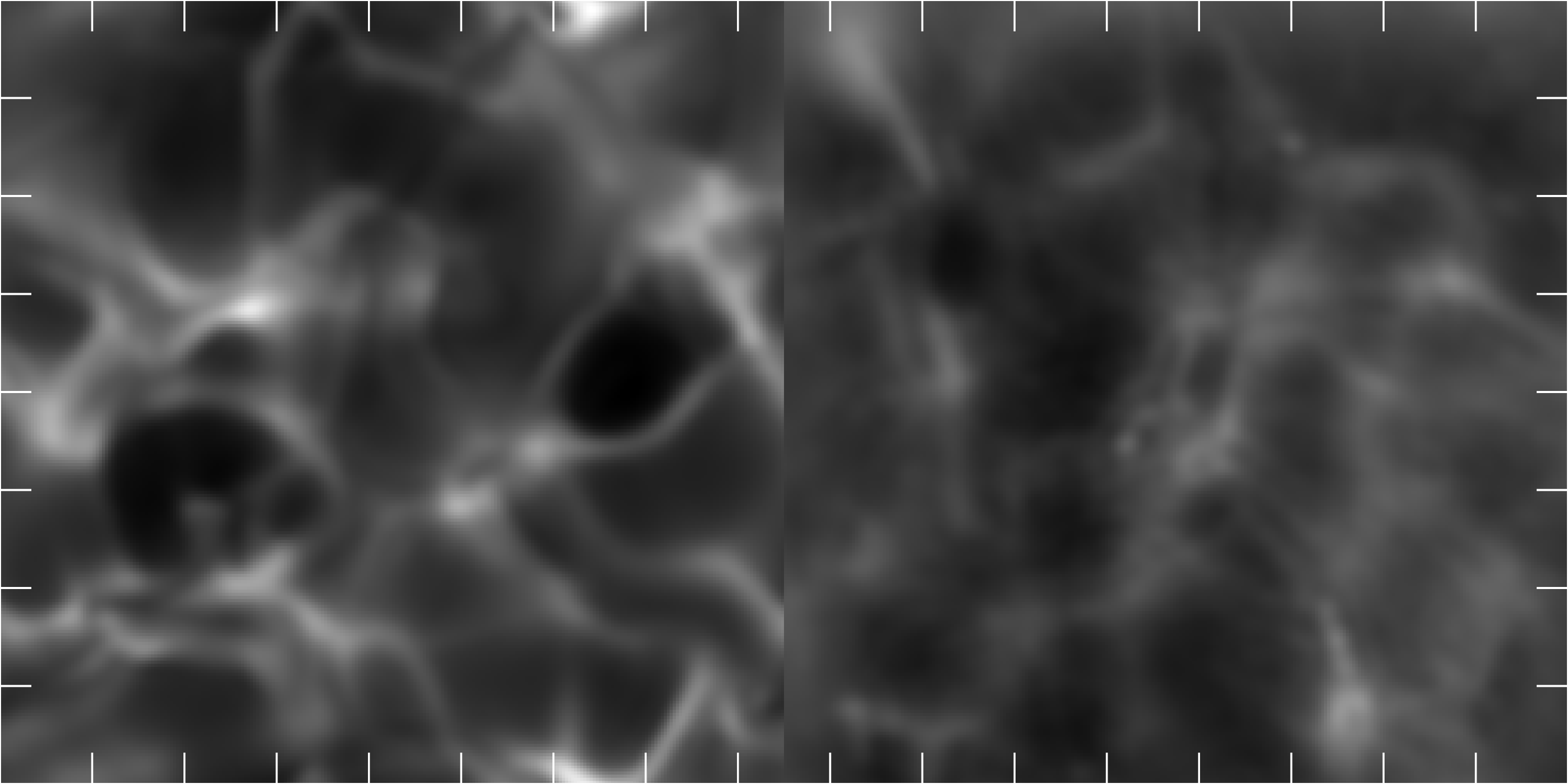}} &   \resizebox{!}{3.3cm}{ \includegraphics{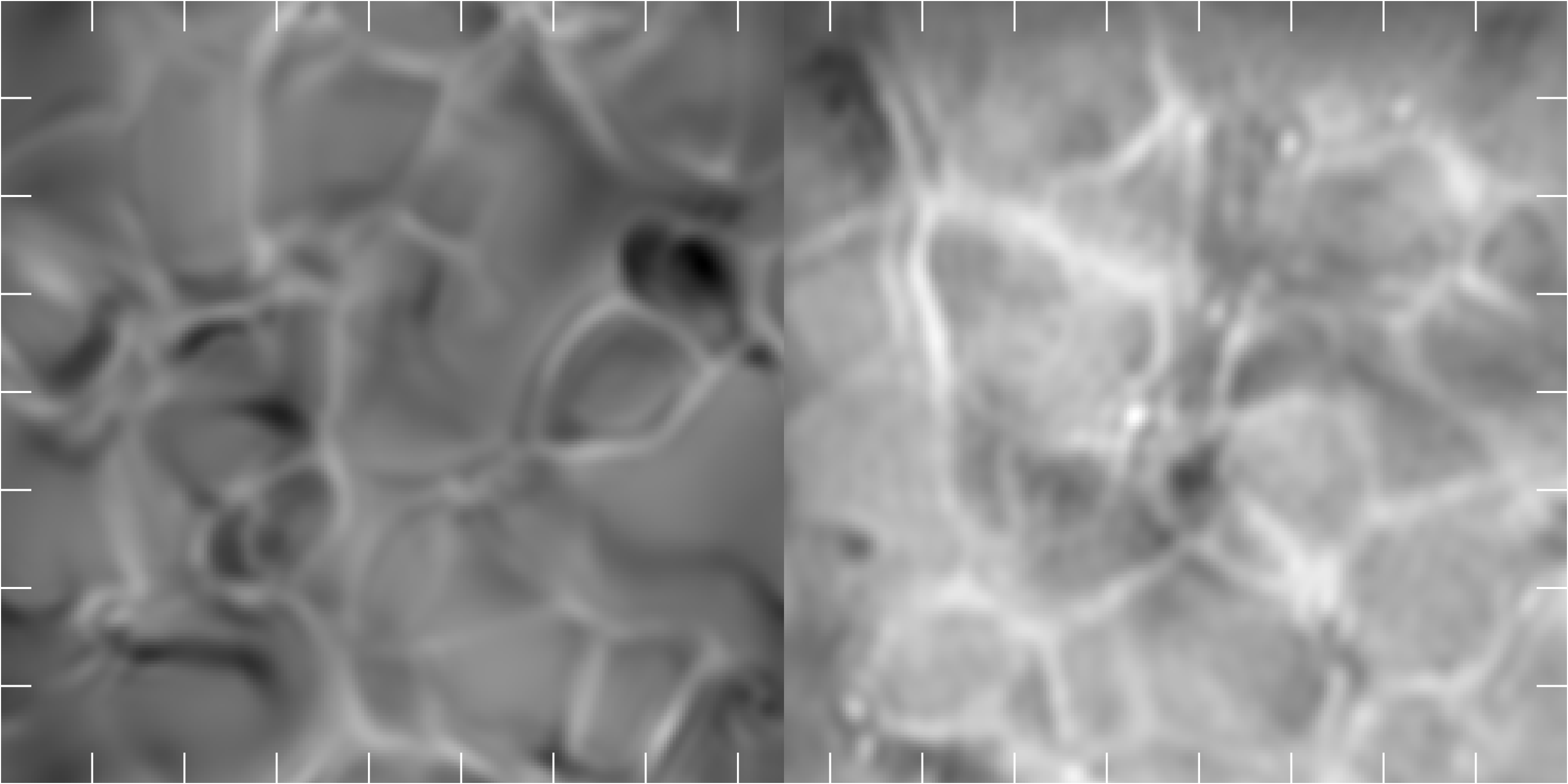}} &  \lab{\centerline{$G_{12}$} 0.54~$\mathrm{g\ cm^{-2}}$} \\ 

              \lab{\centerline{(2)} 0.90~$\mathrm{g\ cm^{-2}}$} &   \resizebox{!}{3.3cm}{ \includegraphics{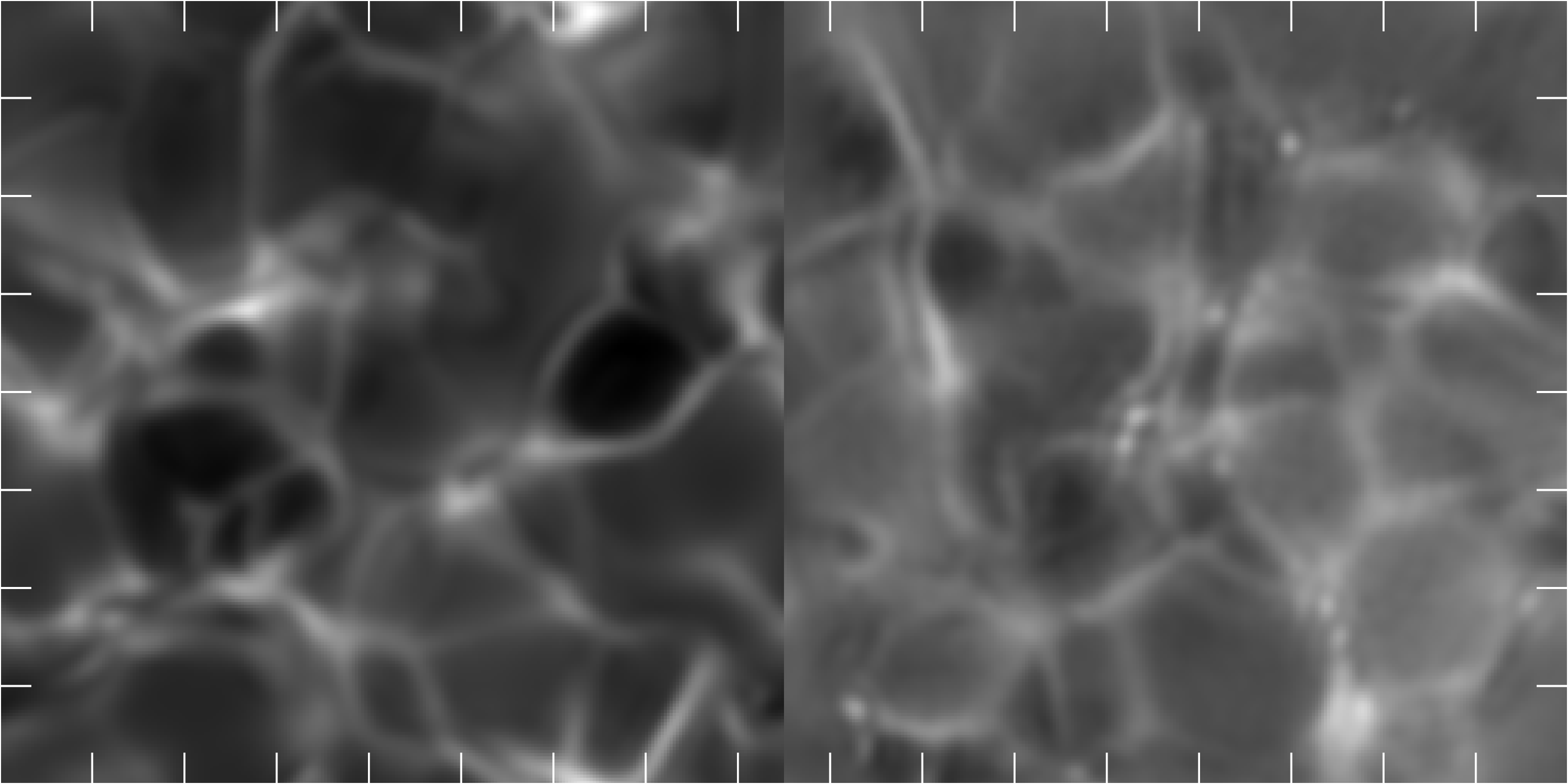}} &  \resizebox{!}{3.3cm}{ \includegraphics{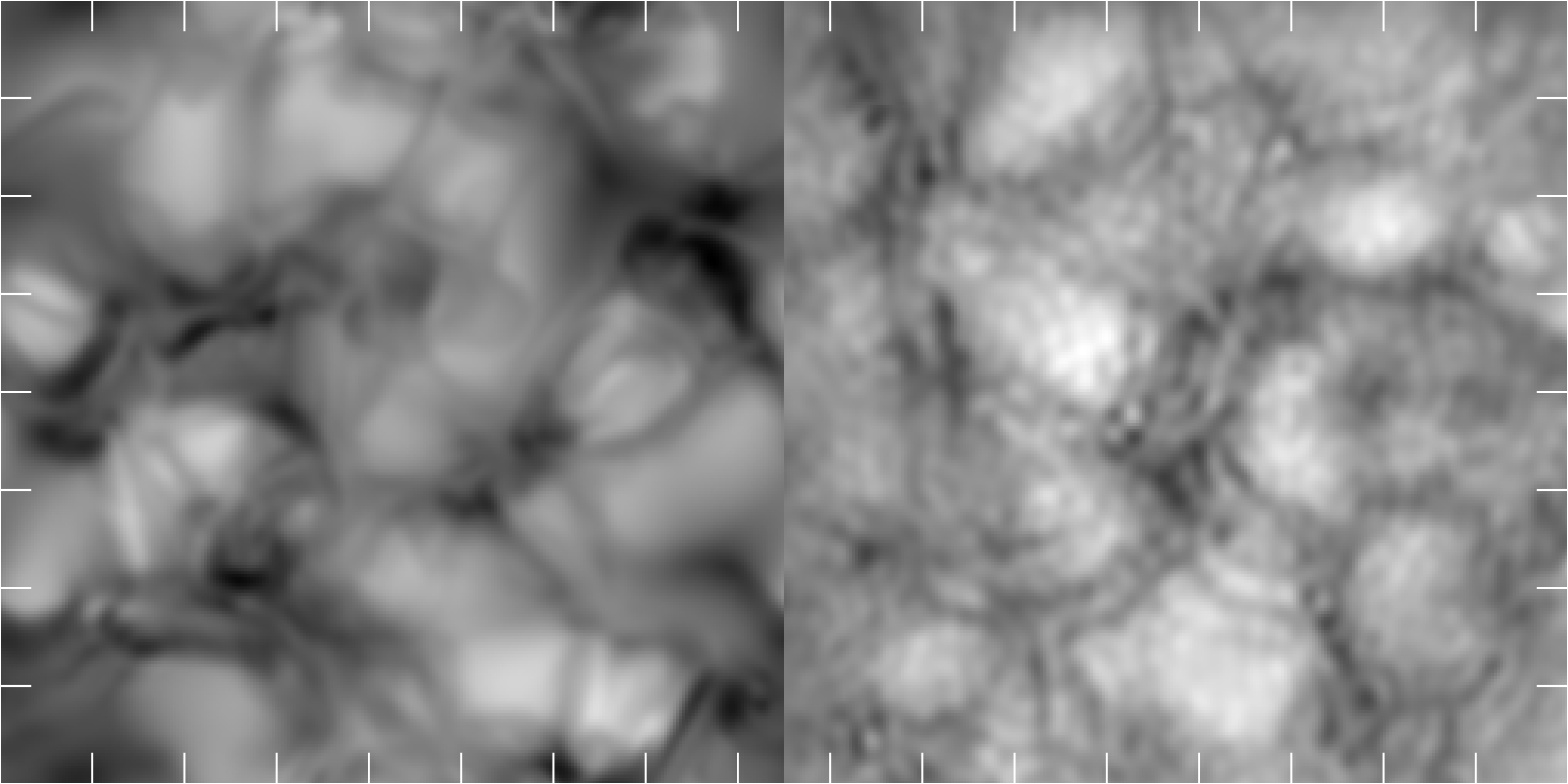}} &  \lab{\centerline{$G_{23}$} 1.15~$\mathrm{g\ cm^{-2}}$}    \\

             \lab{\centerline{(3)} 1.40~$\mathrm{g\ cm^{-2}}$}    &   \resizebox{!}{3.3cm}{  \includegraphics{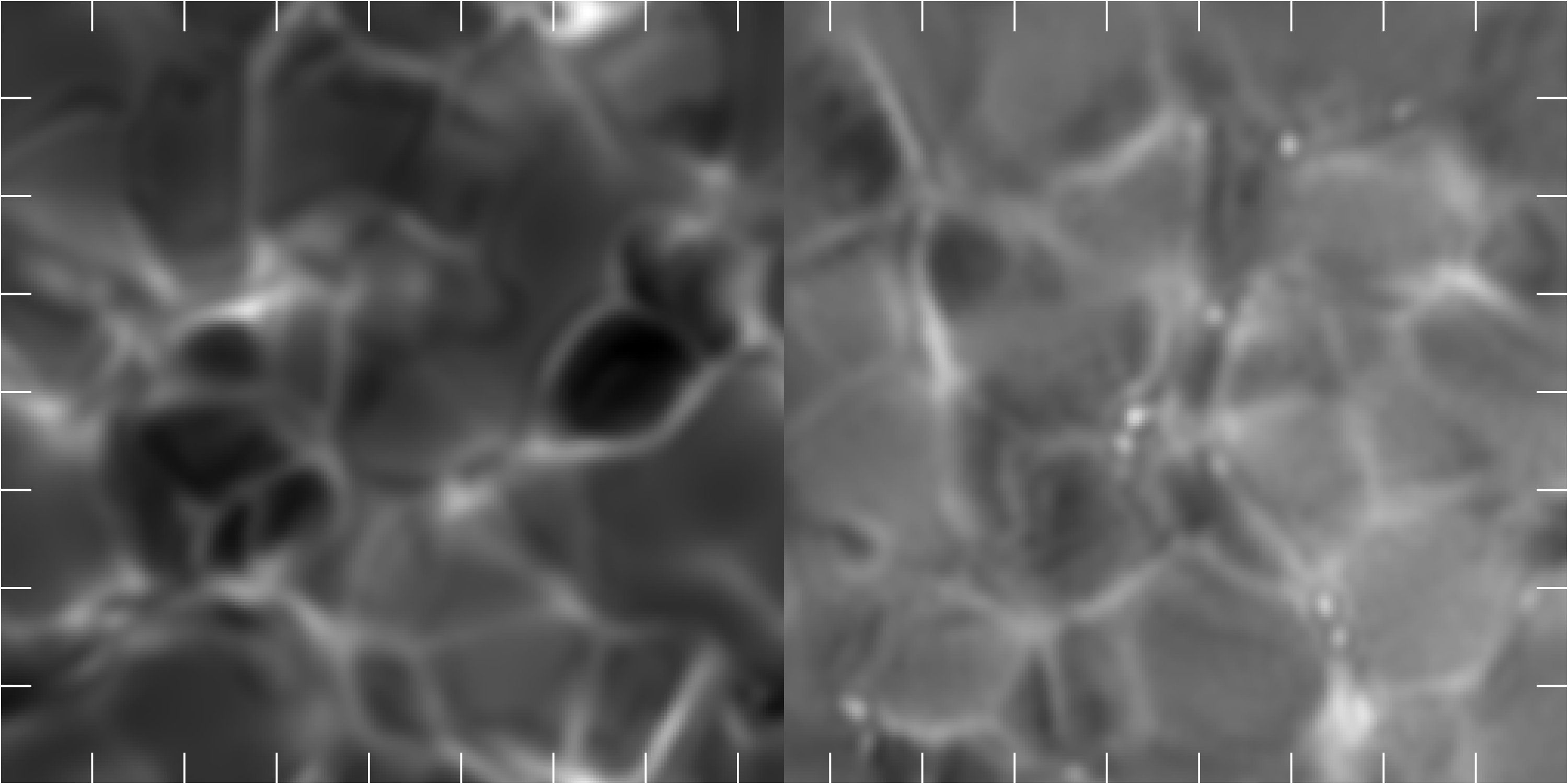}} &   \resizebox{!}{3.3cm}{ \includegraphics{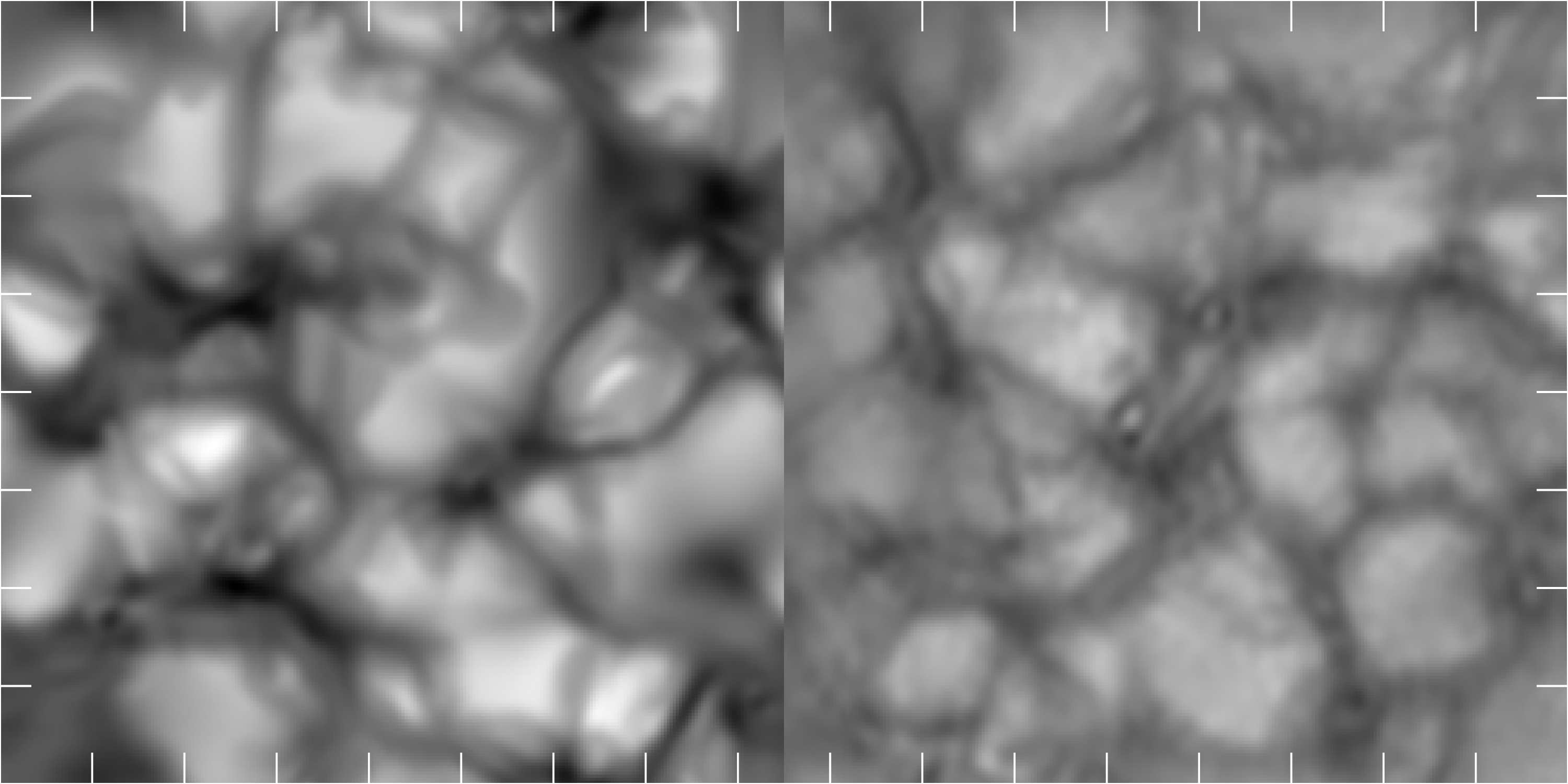}} & \lab{\centerline{$G_{34}$} 1.78~$\mathrm{g\ cm^{-2}}$}\\

             \lab{\centerline{(4)} 2.18~$\mathrm{g\ cm^{-2}}$} &  \resizebox{!}{3.3cm}{ \includegraphics{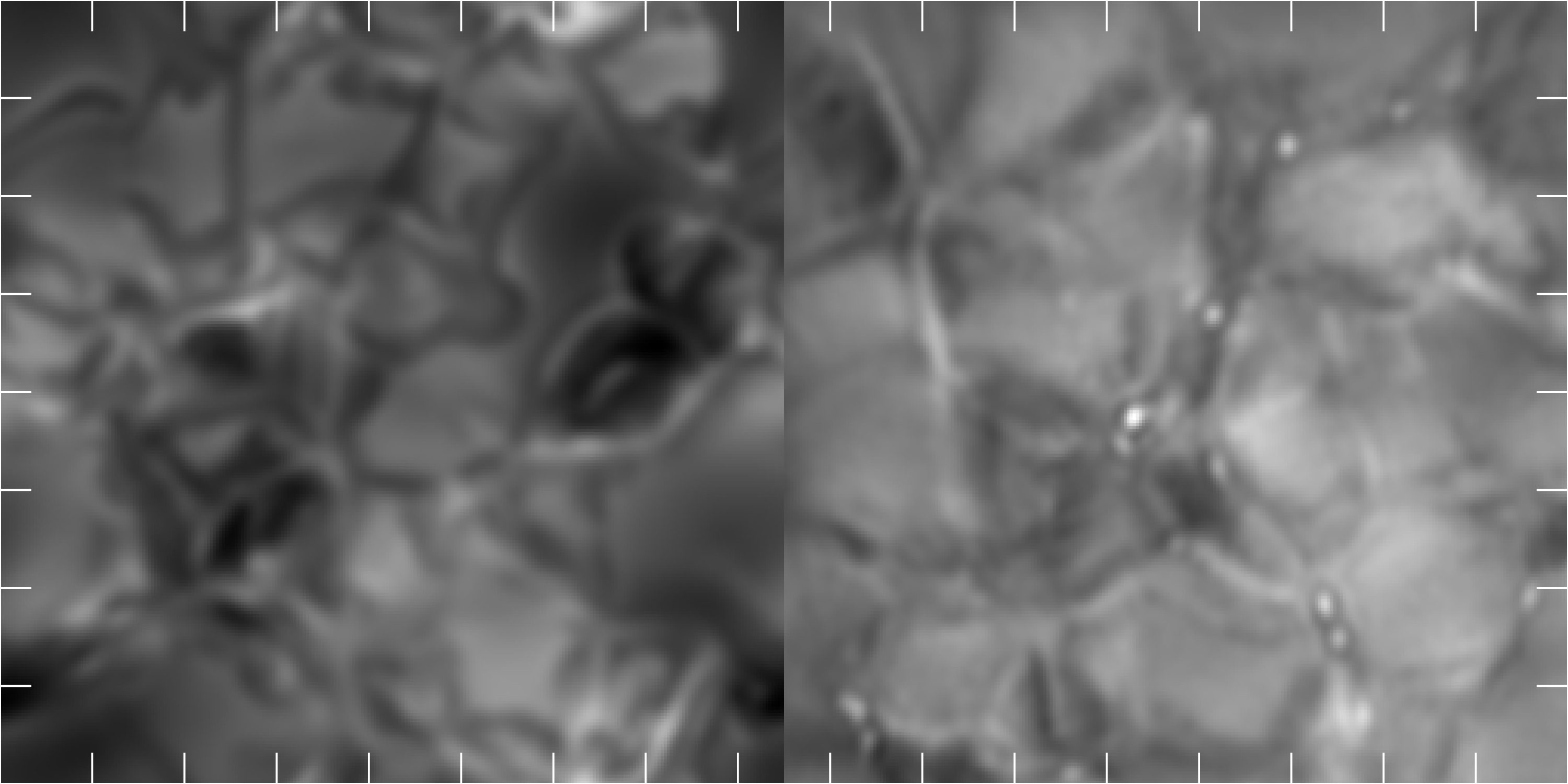}} &  \resizebox{!}{3.3cm}{ \includegraphics{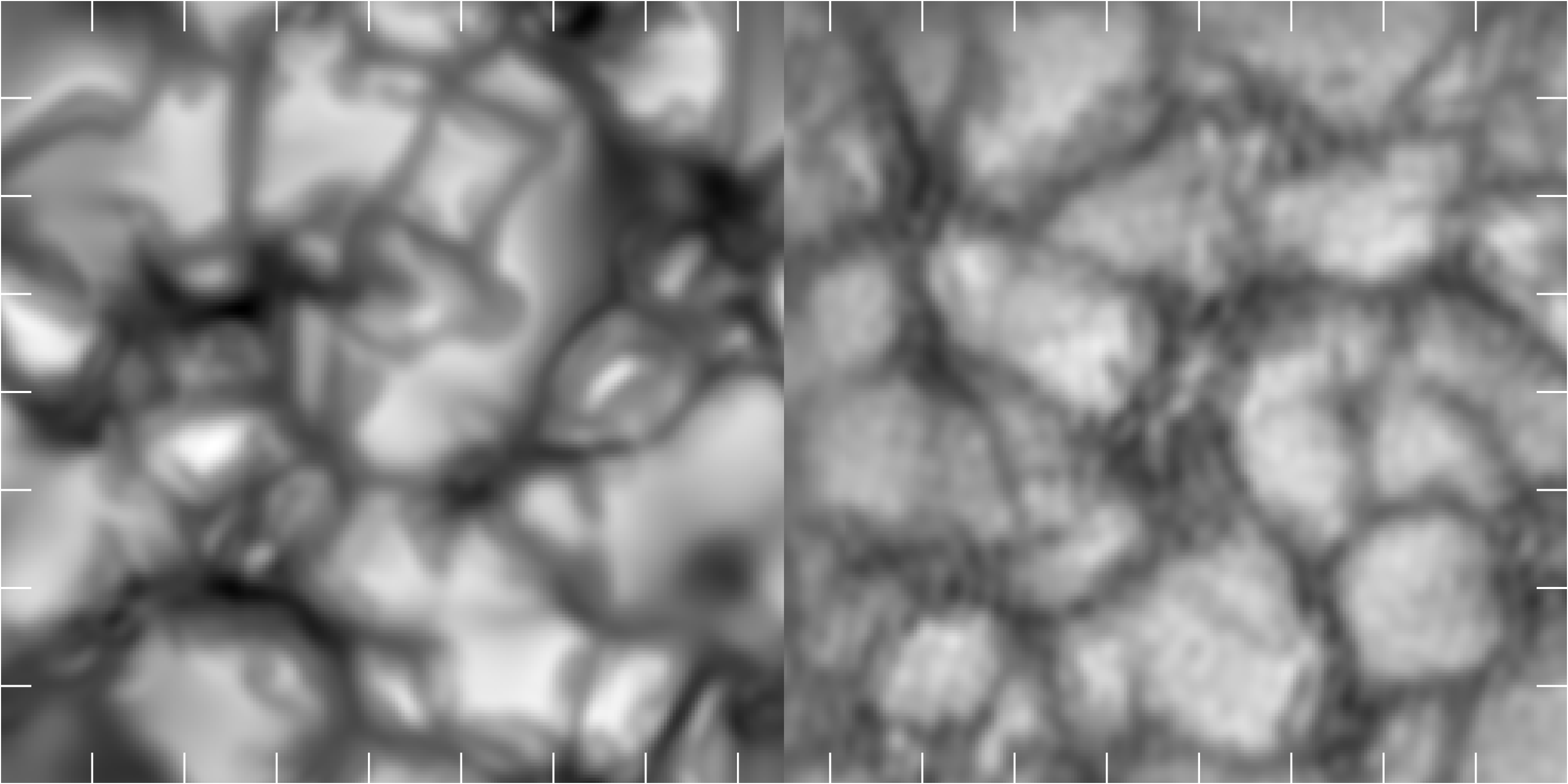}}  &  \lab{\centerline{$G_{45}$} 2.39~$\mathrm{g\ cm^{-2}}$} \\

         \lab{\centerline{(5)} 2.61~$\mathrm{g\ cm^{-2}}$} &  \resizebox{!}{3.3cm}{  \includegraphics{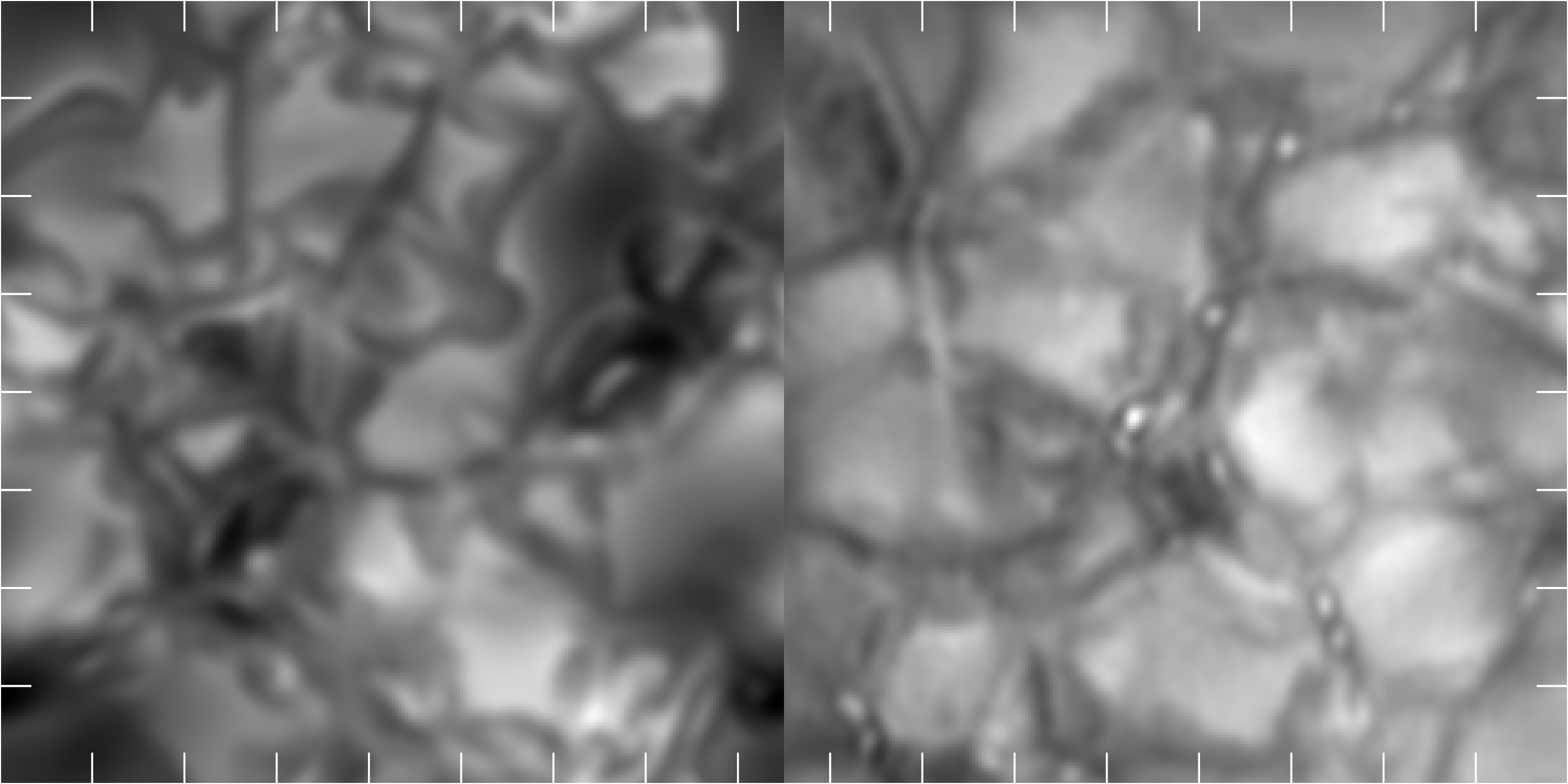}} &     \resizebox{!}{3.3cm}{ \includegraphics{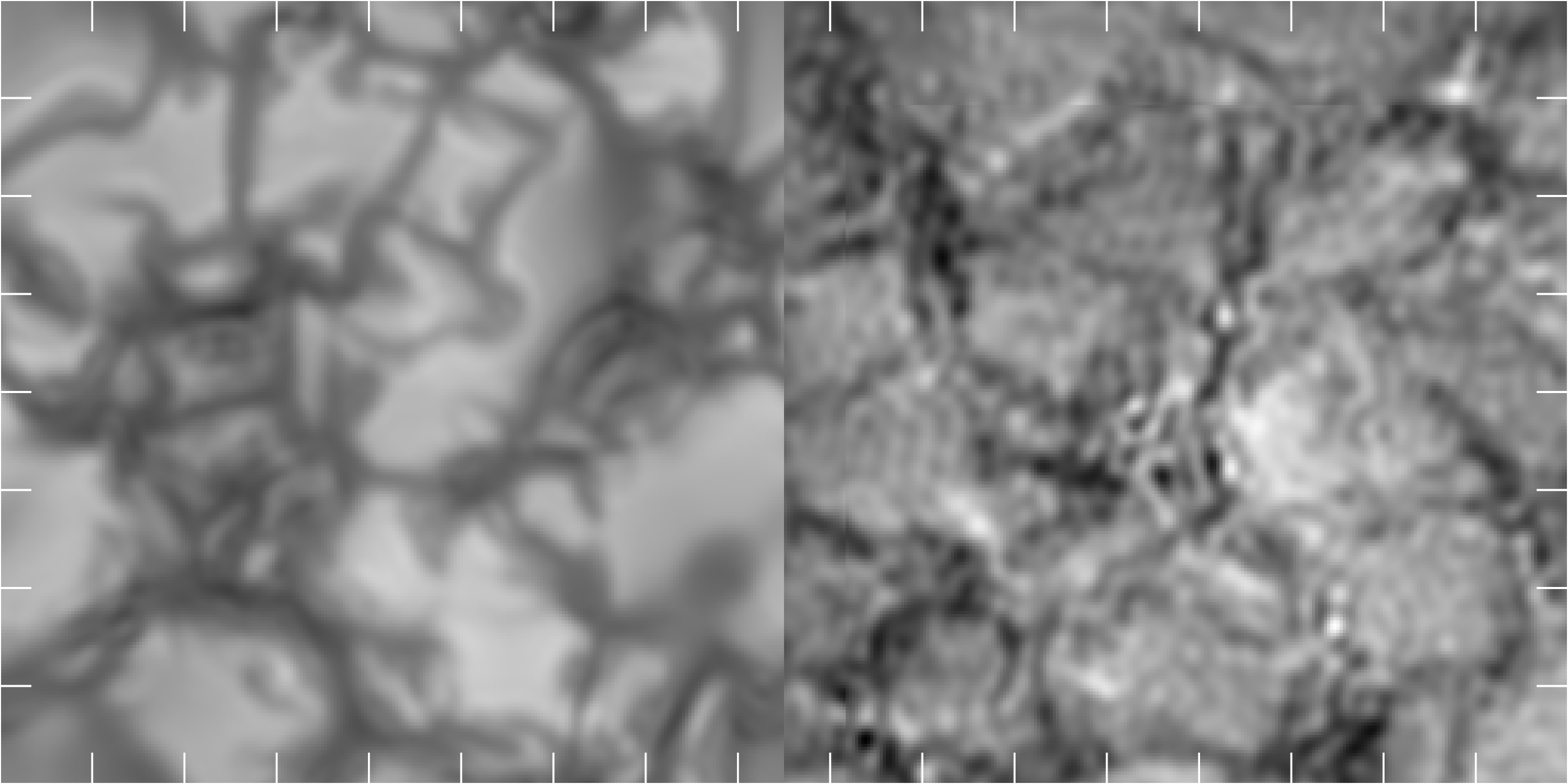}} & \lab{\centerline{$G_{56}$} 2.82~$\mathrm{g\ cm^{-2}}$}\\

       \lab{\centerline{(6)} 3.04~$\mathrm{g\ cm^{-2}}$} &   \resizebox{!}{3.3cm}{ \includegraphics{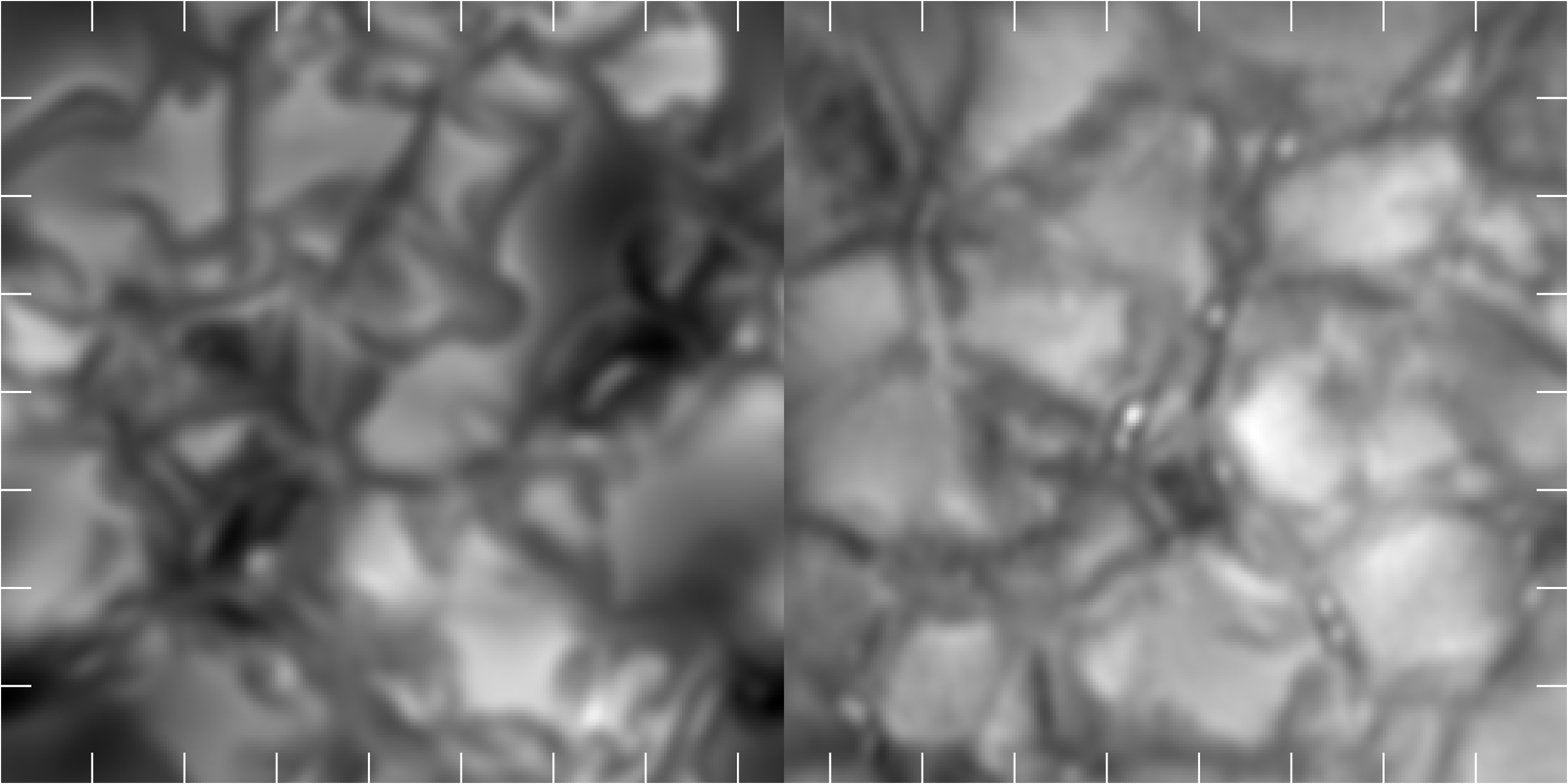}} &   \resizebox{!}{3.3cm}{ \includegraphics{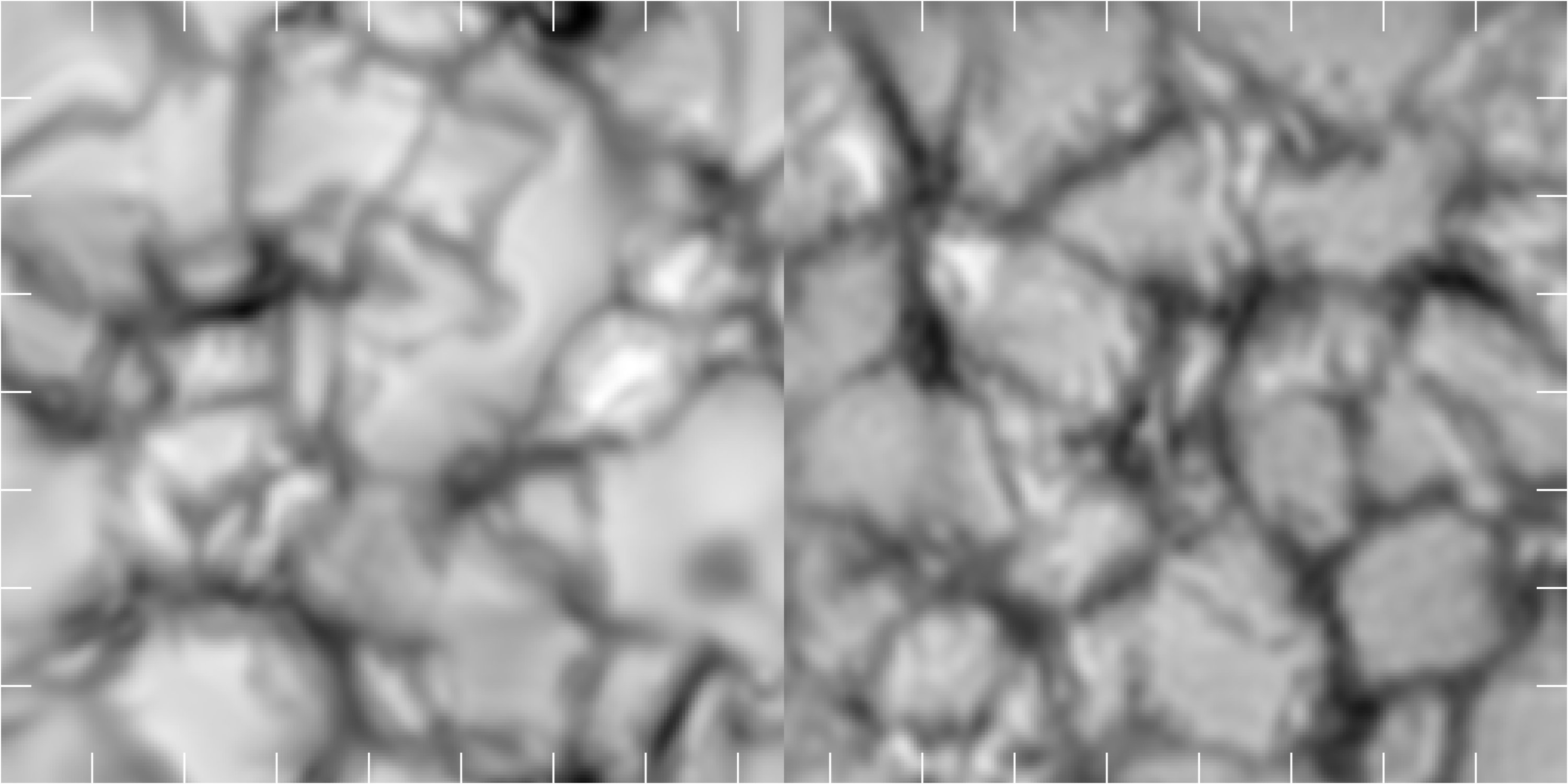}}  & \lab{\centerline{$G_{57}$} 3.09~$\mathrm{g\ cm^{-2}}$} \\
       \lab{\centerline{(7)} 3.56~$\mathrm{g\ cm^{-2}}$}  &   \resizebox{!}{3.3cm}{ \includegraphics{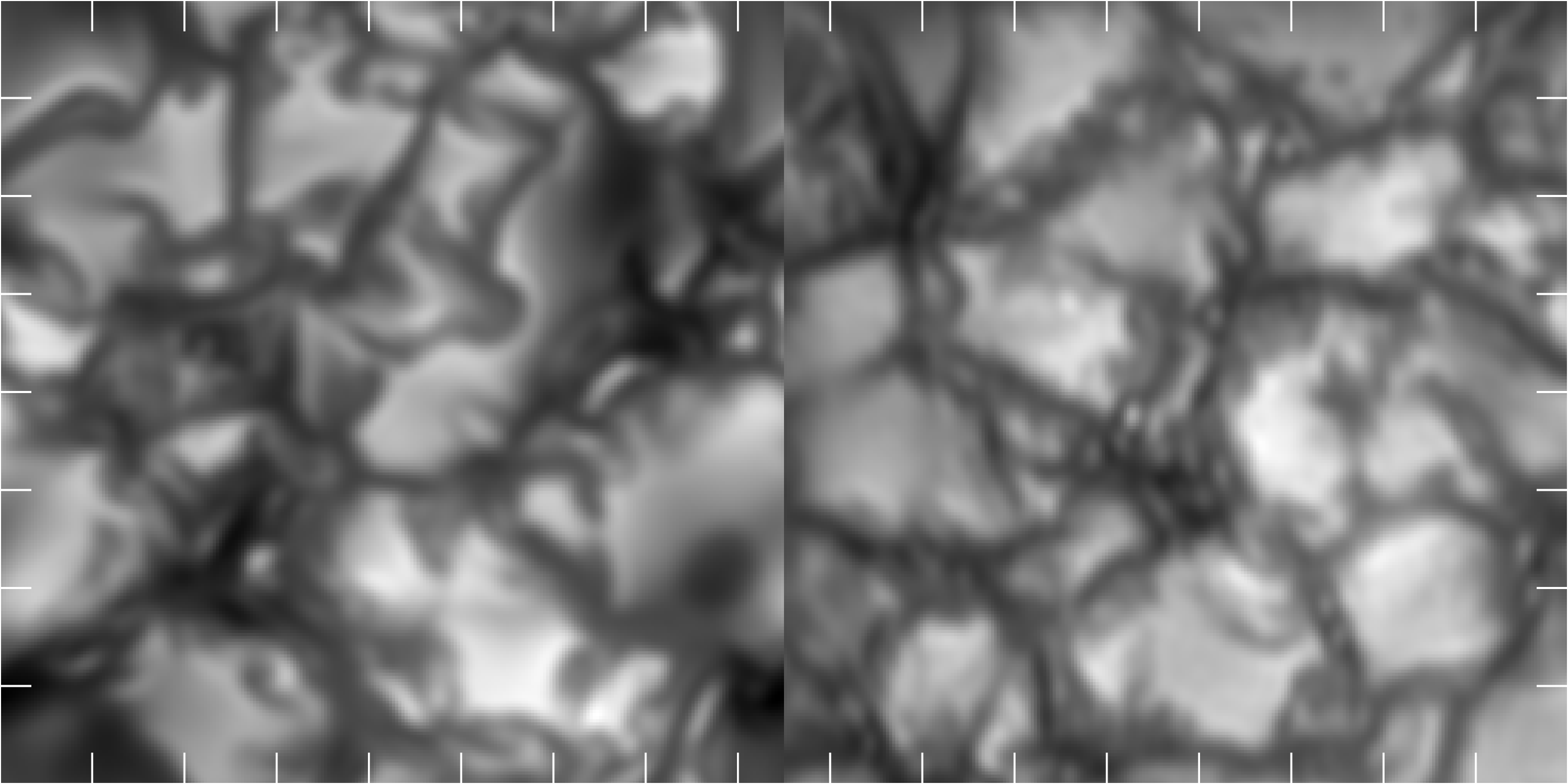}} & \resizebox{!}{3.3cm}{ \includegraphics{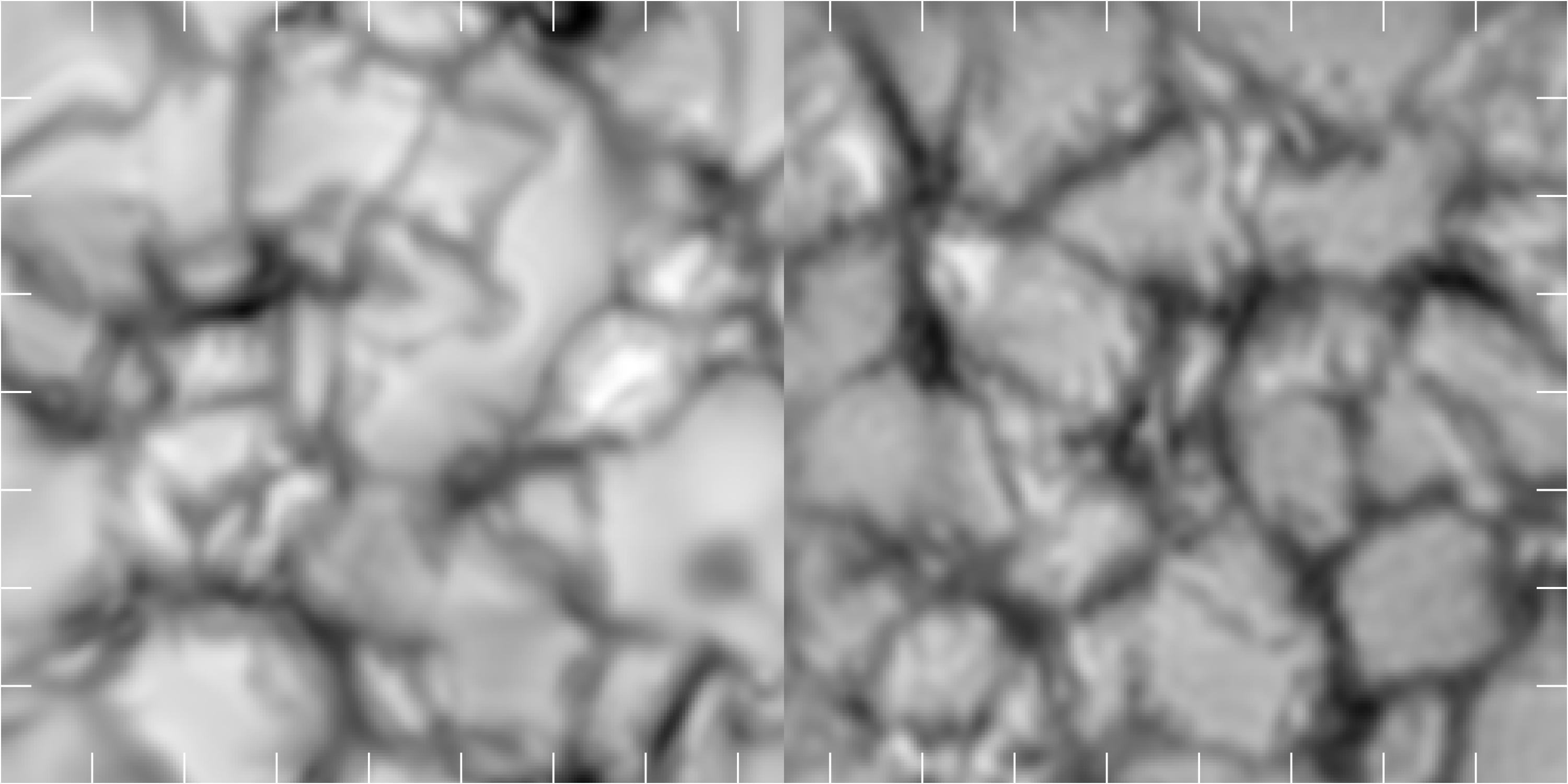}}   & \lab{\centerline{$G_{67}$} 3.30~$\mathrm{g\ cm^{-2}}$}

    \end{tabular}
  \caption{\footnotesize Relatively quiet Sun patch (upper left corner of Fig.~\ref{wideband}) compared with the synthetic images from the HD snapshot. From left to right: intensity from synthetics, intensity from observations, gradient maps from synthetics and gradient maps from observations. The depths in terms of column mass shown on the left side are the average depth from all the pixels when computed as in Sect.~\ref{sub:depth}. Each tick mark corresponds to one arcsec. Each pair of synthetic and observational images are scaled in the same way.}   
 
      \label{intonly}
\end{figure*}

\section{Comparison of observations and simulations}
\label{sec:gradients}

\subsection{Quiet Sun and HD synthetic data}
\label{HDcomp}

In Fig.~\ref{intonly}, a relatively quiet area from the observations is compared with the HD simulation synthetic data for all the heights sampled. Since intensity and temperature maps look essentially the same, we show only intensity and temperature gradient maps. All the maps, at each height, have the same scale pairwise (each pair of synthetic data and observations). The intensity maps for the synthetics shown in the left column are not calibrated to HolMul, so that the difference in overall intensity is visible. In the right half of Fig.~\ref{intonly}, temperature gradient maps computed as described in the previous sections are shown. The similarity between observations and simulations is striking.

The deepest layer (position (7) or 3.56~$\mathrm{g\ cm^{-2}}$) shows an excellent match in both morphology and overall intensity level between simulations and synthetics. However, what is remarkable is the similarity in the deep gradient maps ($G_{67}$ at 3.30~$\mathrm{g\ cm^{-2}}$ and $G_{57}$ at 3.09~$\mathrm{g\ cm^{-2}}$ ) as (7) is the filter position affected by the largest errors. Despite their large size, such errors are systematic and structure dependent, as seen in the previous section. Thus they affect structures present in both synthetics and observations in the same way. At this depth, both synthetic and observational temperature gradient maps show the interior of granules nearly uniformly as high-gradient structures and intergranular lanes as low-gradient structures. This is a pattern that holds up to the gradient maps at 1.15~$\mathrm{g\ cm}^{-2}$, with the difference between granules and intergranular lanes gradually decreasing with height.

The observational gradient map at position $G_{56}$ is clearly noisier than the rest and appears very different from that of the simulations. Position (6) is polluted by the strong \ion{Al}{i} blend, thus presenting mixed information from the different heights. The presence of the blend shifts the formation height higher, effectively making it closer to position (5) than to position (7). This reduces the baseline from which gradients are computed, causing noise to become more visible. 

Looking at the intensity maps for positions (6) to (4), at 0.90~$\mathrm{g\ cm^{-2}}$ to 2.18~$\mathrm{g\ cm^{-2}}$, one sees that the transition from granulation to inverse granulation becomes apparent at a greater height in the simulations. 

Comparing the intensity maps in positions (2) and (1), we see that the pattern is nearly the same for both maps in the simulation data. We note that the simulation does not aim at reproducing the chromosphere, so this is not surprising. However, for the observations, this pattern is also nearly the same but fuzzier for position (1). This is very suggestive of radiative scattering effects that are not captured by the 1.5 D radiative transfer used to produce the synthetic images. Scattering in the upper photospheric layers as observed in \ion{Ca}{ii}~K was discussed by \cite{1999ASPC..184..181R} in the context of the observed fuzziness in network grains when compared with other diagnostics. While these layers are known to be highly dynamic, waves and heating mechanisms not present in the lower layers (such as the ones identified by \cite{2010MmSAI..81..582C} and discussed in \cite{2012RSPTA.370.3129R}) seem to affect mainly the shape of the intensity pattern. What we see in the area shown, however, is a constant pattern with an increase in ``fuzziness''. 

In the observational intensity maps, small bright points are visible that appear to have nearly the same size for all heights with the exception of the very diffuse line core position. Carefully inspecting the WB image, one can argue that the small bright points that are visible in the upper layers, including some as small as 5 pixels as measured from minimum to minimum, are about one pixel (0\farcs034) wider than in the WB when looking at position (4) or higher in the atmosphere. This hints at expansion with height. However, this is close to our resolution limit, and it might be that these points are not resolved.

Figure~\ref{intonly} illustrates that the highest possible spatial resolution is necessary to capture small-scale dynamics at these heights, even in the absence of magnetic fields. The morphological match between simulated data and observations in both intensity and temperature gradient maps strongly suggests that the physics of granulation is well captured by the HD simulation, but differences exist between the synthetic images and observations when looking at the progression with height.

\subsection{An active area and MHD synthetic data}
\label{compareMHD}
In this section, we focus on interpreting Fig.~\ref{tmaps}, which shows a selected region of 10\arcsec~($\sim$7.1~Mm) from the synthetic data produced from the MHD simulation and a similar active area from the observations. This figure is built in the same way as Fig.~\ref{intonly}. However, due to space constraints and in order to show a larger image area that accommodates both pores and bright points, we cut out two rows. We choose to remove the images for position (6) and the gradients maps $G_{67}$ and $G_{56}$ due to the pollution of position (6) from the \ion{Al}{i} blend as stated in Sect.~\ref{HDcomp}. The intensity map from position (1) is also removed because it is dominated by chromospheric features not present in the simulation. Both synthetics and observations present pores, and there are no point-like isolated bright features (like the ones visible in Fig.~\ref{intonly}) in either the observed or the synthetic data in the selected areas. Even though the simulation atmosphere is much colder than the observations in the upper layers, which precludes detailed quantitative comparison, the morphological similarity between simulation and observations is impressive both in intensity and temperature gradient maps.

For the deepest layer ($G_{57}$ at 3.09~$\mathrm{g\ cm}^{-2}$) there is a region of strongly reduced temperature gradient at and surrounding the pores as well as bright points for both observations and synthetics (visible as a sprawling dark area in the gradient maps). This region is identifiable in the gradient maps up to 1.15~$\mathrm{g\ cm}^{-2}$ ($G_{23}$ ), but between 2.39~$\mathrm{g\ cm}^{-2}$ and 1.15~$\mathrm{g\ cm}^{-2}$ ($G_{45}$ and $G_{23}$) small higher gradient structures (visible as bright filigree structures in the gradient maps) start rimming the outer border of this low-gradient area. These high-gradient structures expand and become stronger with height (relative to their surroundings). At the topmost gradient map ($G_{12}$ at 0.54~$\mathrm{g\ cm}^{-2}$)  for the synthetic data and at 1.15~$\mathrm{g\ cm}^{-2}$ ($G_{23}$) for the observations, the expansion of these high-gradient structures leads to a near reversion of the pattern seen for the deepest layers ($G_{57}$), with bright points showing higher temperature gradient than granules. Intergranular lanes are not distinguishable in the gradient maps at these heights (as was the case for the field-free HD simulation) for either synthetics or observations. The central part of the pores displays low gradient at all heights.

In the deep layers, bright points have lower temperature gradient compared with their surroundings. A likely explanation for this is hot-wall radiation heating the mid-photospheric optical depths inside fluxtubes, causing a ``flatter gradient'' as mentioned in \cite{2010rasp.book..163R} when reproducing the hot-wall cartoon from \cite{1976SoPh...50..269S}.

Nearly all bright-point chains (``ribbons'' and ``flowers'' as coined by \cite{2004A&A...428..613B}) expand with height both in the synthetic and the observational intensity maps of Fig.~\ref{tmaps}. This seems to be a different behavior than that seen in the isolated bright points mentioned in Sect.~\ref{HDcomp}. The latter expand slightly, if at all, in the images, maybe because they are unresolved. 

A ringing artifact can be seen in the lower left part of the bottom gradient map for the synthetic data (at $x=2\arcsec,y=3\arcsec$ with origin in the lower left corner). This is purely due to the noise filtering and to the extreme sharpness of the feature. Differentiating data enhances such artifacts. To a smaller degree, such ringing is also visible in the observed gradient maps but is hard to separate from the noise. The synthetic data are useful in drawing attention to such limitations of the data processing. 

The simulations replicate the observed chains of bright points and their expansion with height in intensity as well as pores. They also replicate the complex small-scale structure visible in the observational temperature gradient maps. This degree of morphological similarity between simulated data and observational data for both the MHD and HD cases makes it likely that both simulations capture the essential physics of the different observed structures.

\begin{figure*}[!hl]

  \def\lab#1{\begin{minipage}[b]{14mm}
      #1\vspace{15mm}
    \end{minipage}}
  \def\tile#1{\resizebox{!}{3.3cm}{ \includegraphics{#1}}}
\begin{tabular}{c@{\hspace{2mm}}c@{\hspace{1mm}}c@{\hspace{2mm}}c}
    \hspace*{-7mm}   &   {\hspace{2mm}}Synthetic Intensity {\hspace{6mm}} Observational Intensity &  {\hspace{2mm}}Synthetic Gradient {\hspace{6mm}} Observational Gradient   &   \\

 \hspace*{-7mm}
   \lab{\centerline{(2)}  0.90~$\mathrm{g\ cm^{-2}}$} &   \resizebox{!}{4.0cm}{  \includegraphics{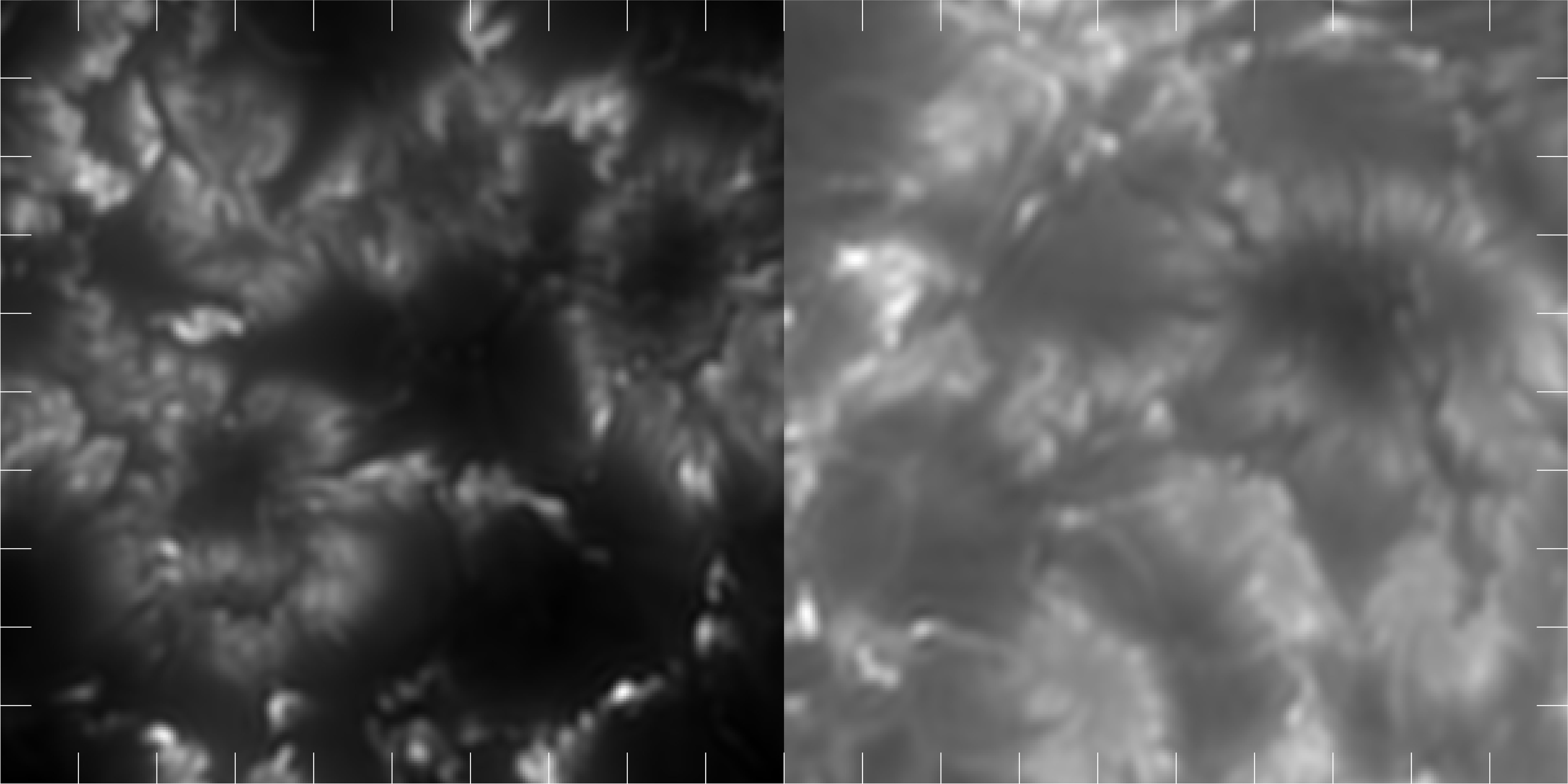}} &  \resizebox{!}{4.0cm}{ \includegraphics{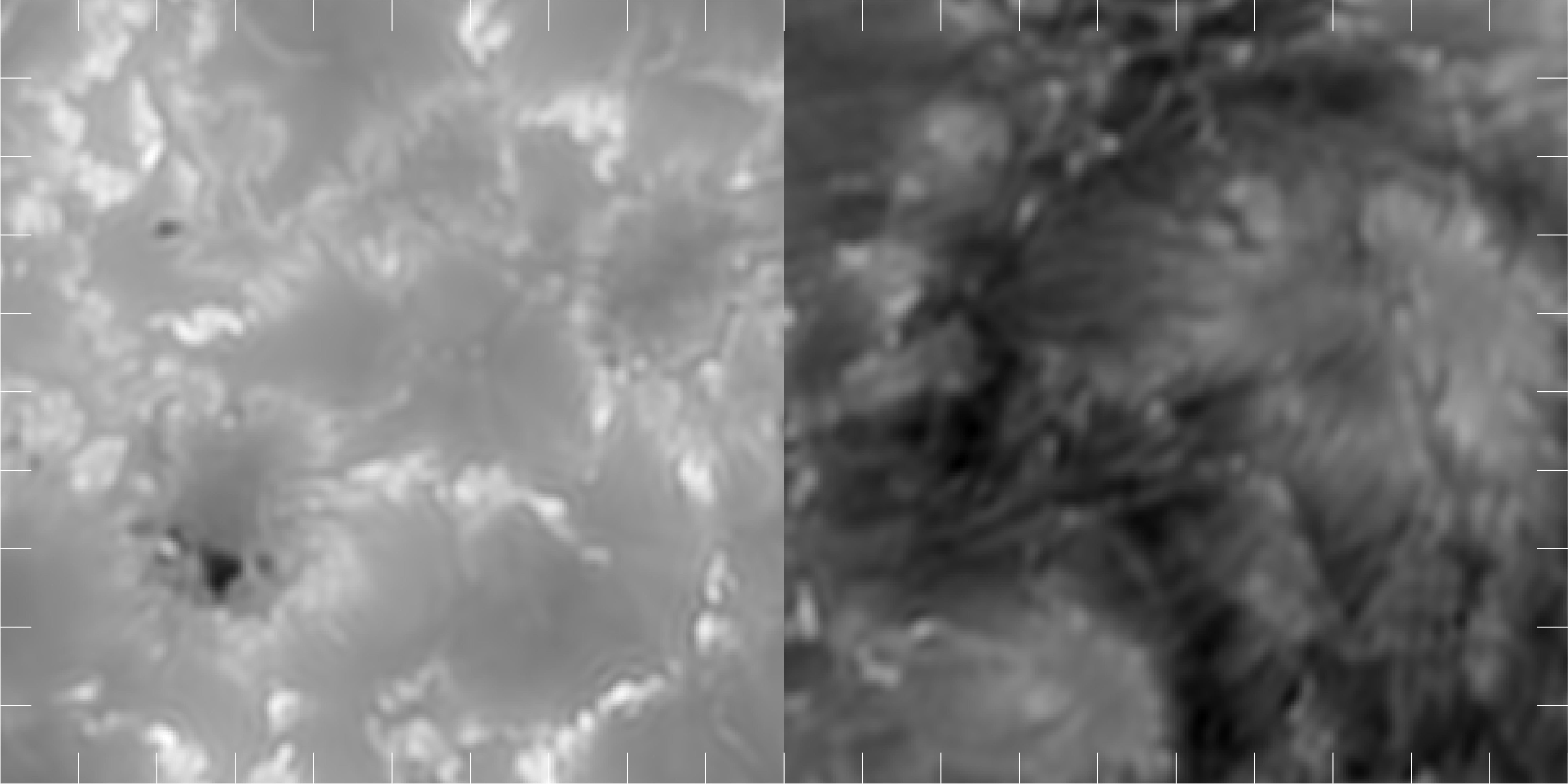}} &  \lab{\centerline{$G_{12}$} 0.54~$\mathrm{g\ cm^{-2}}$} \\ 

\hspace*{-7mm}
      \lab{\centerline{(3)} 1.40~$\mathrm{g\ cm^{-2}}$}   &  \resizebox{!}{4.0cm}{  \includegraphics{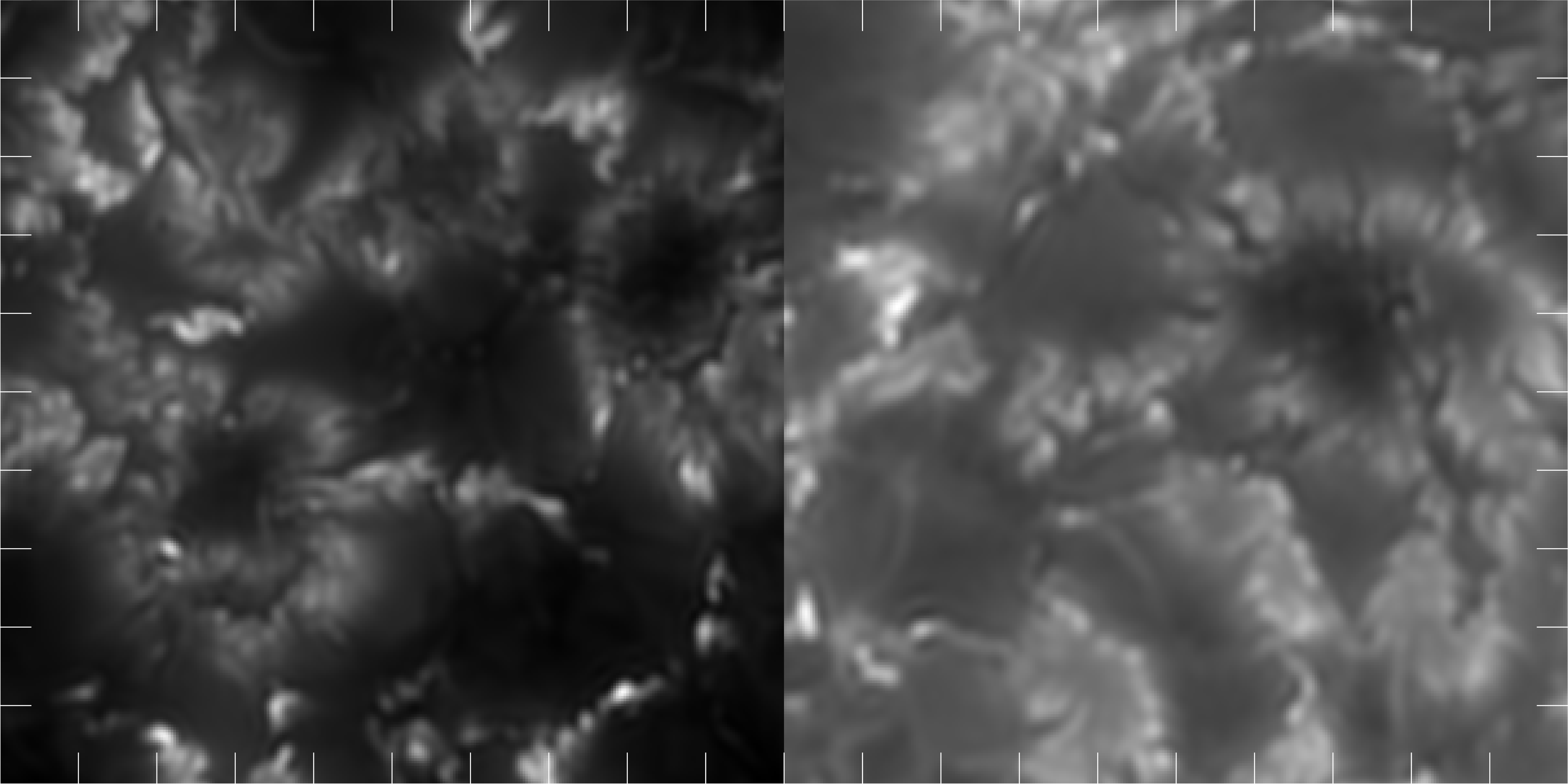}}     &   \resizebox{!}{4.0cm}{  \includegraphics{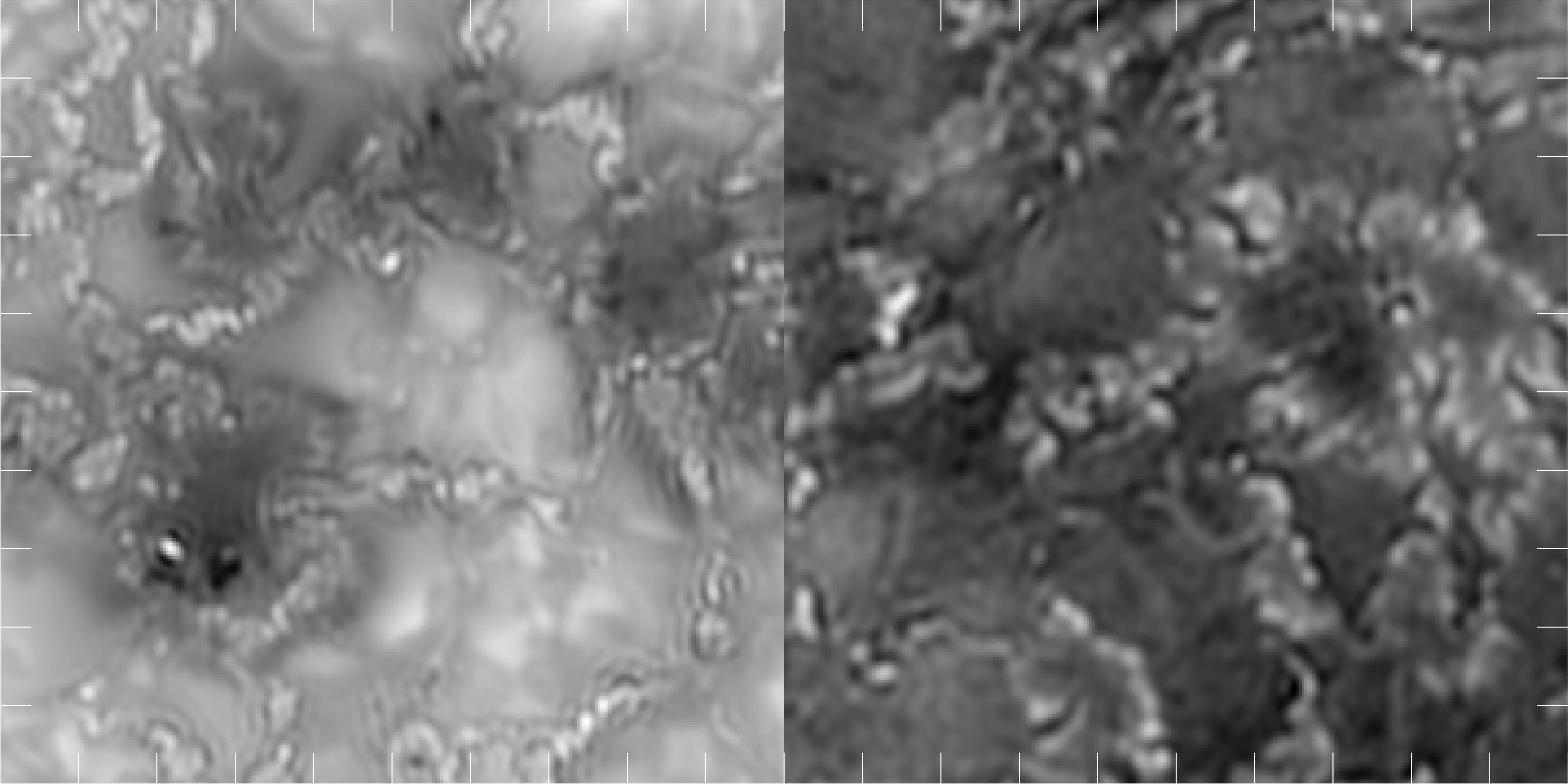}}   &  \lab{\centerline{$G_{23}$} 1.15~$\mathrm{g\ cm^{-2}}$} \\ 

\hspace*{-7mm}
  \lab{\centerline{(4)} 2.18~$\mathrm{g\ cm^{-2}}$} &   \resizebox{!}{4.0cm}{  \includegraphics{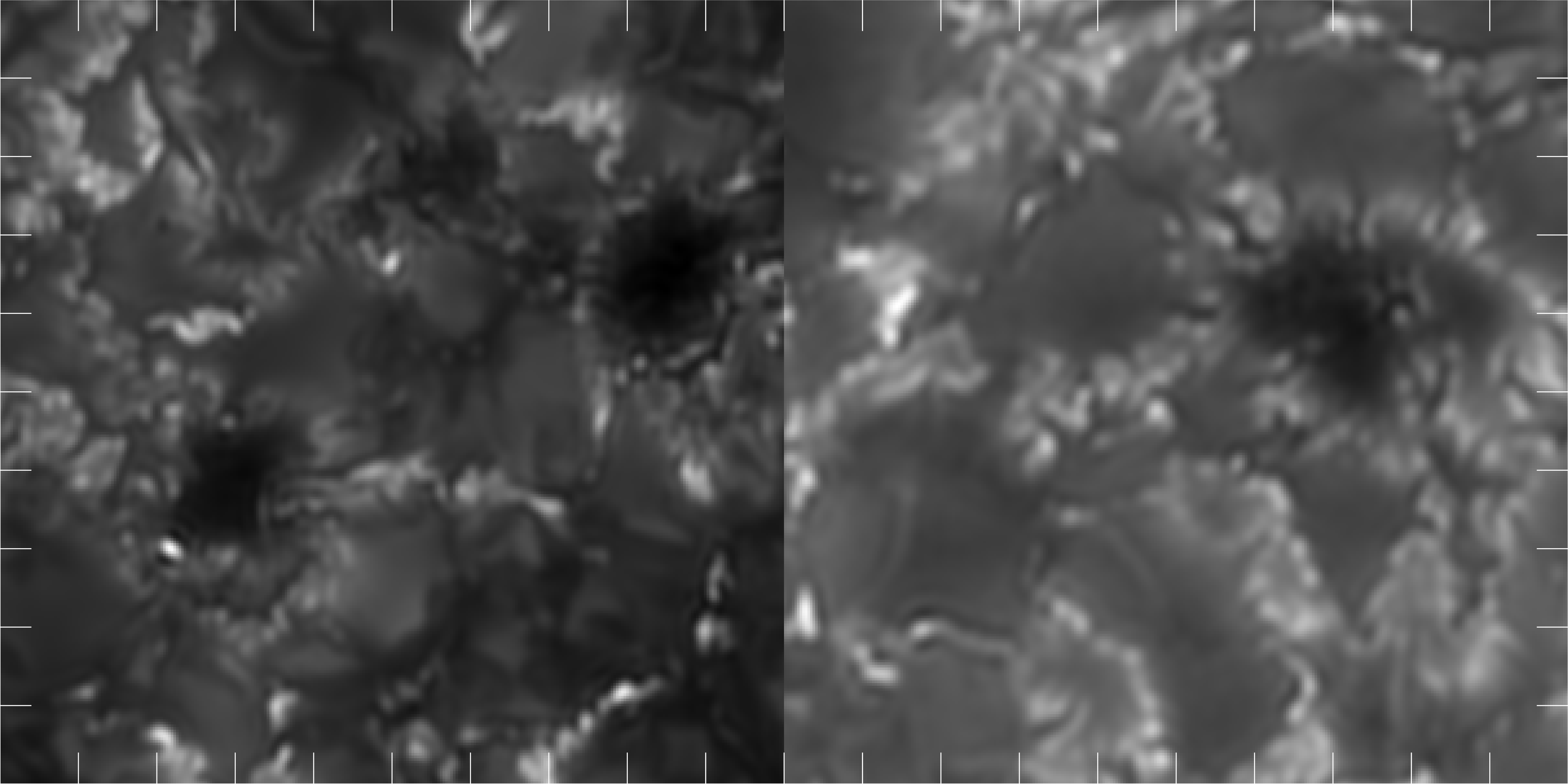}}     &   \resizebox{!}{4.0cm}{ \includegraphics{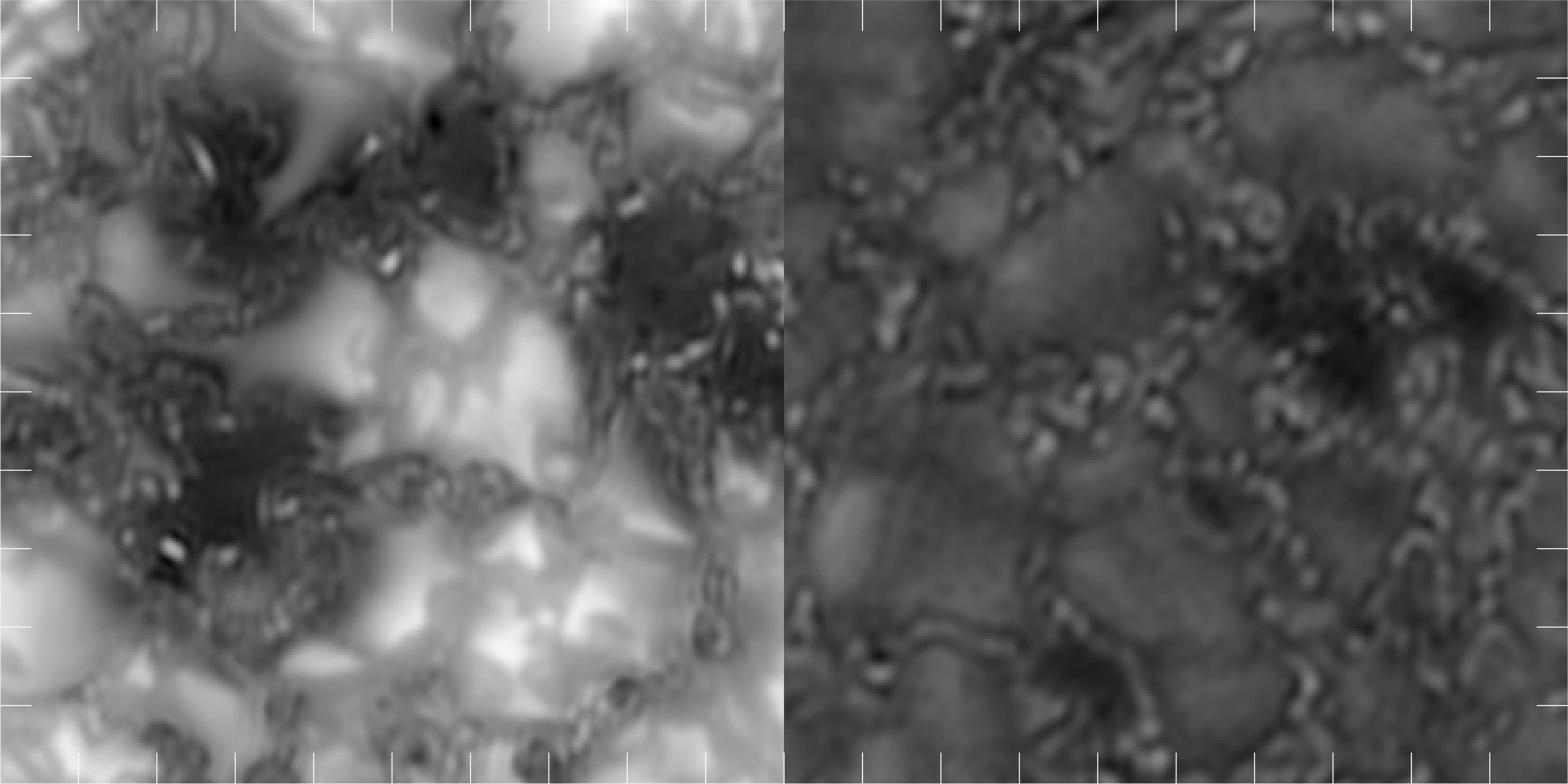}} & \lab{\centerline{$G_{34}$} 1.78~$\mathrm{g\ cm^{-2}}$} \\ 

\hspace*{-7mm}
  \lab{\centerline{(5)} 2.61~$\mathrm{g\ cm^{-2}}$} &   \resizebox{!}{4.0cm}{  \includegraphics{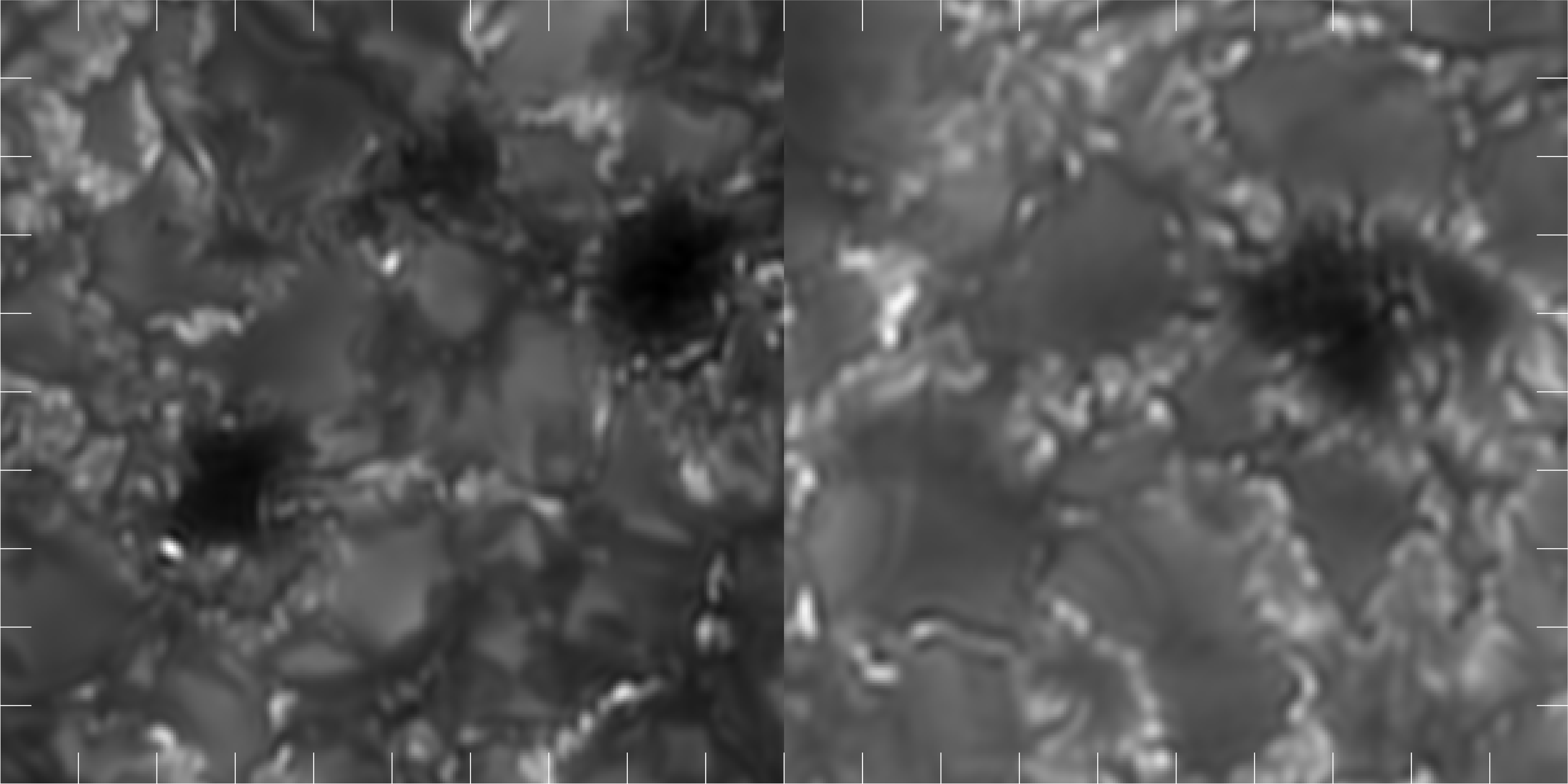}}     &   \resizebox{!}{4.0cm}{ \includegraphics{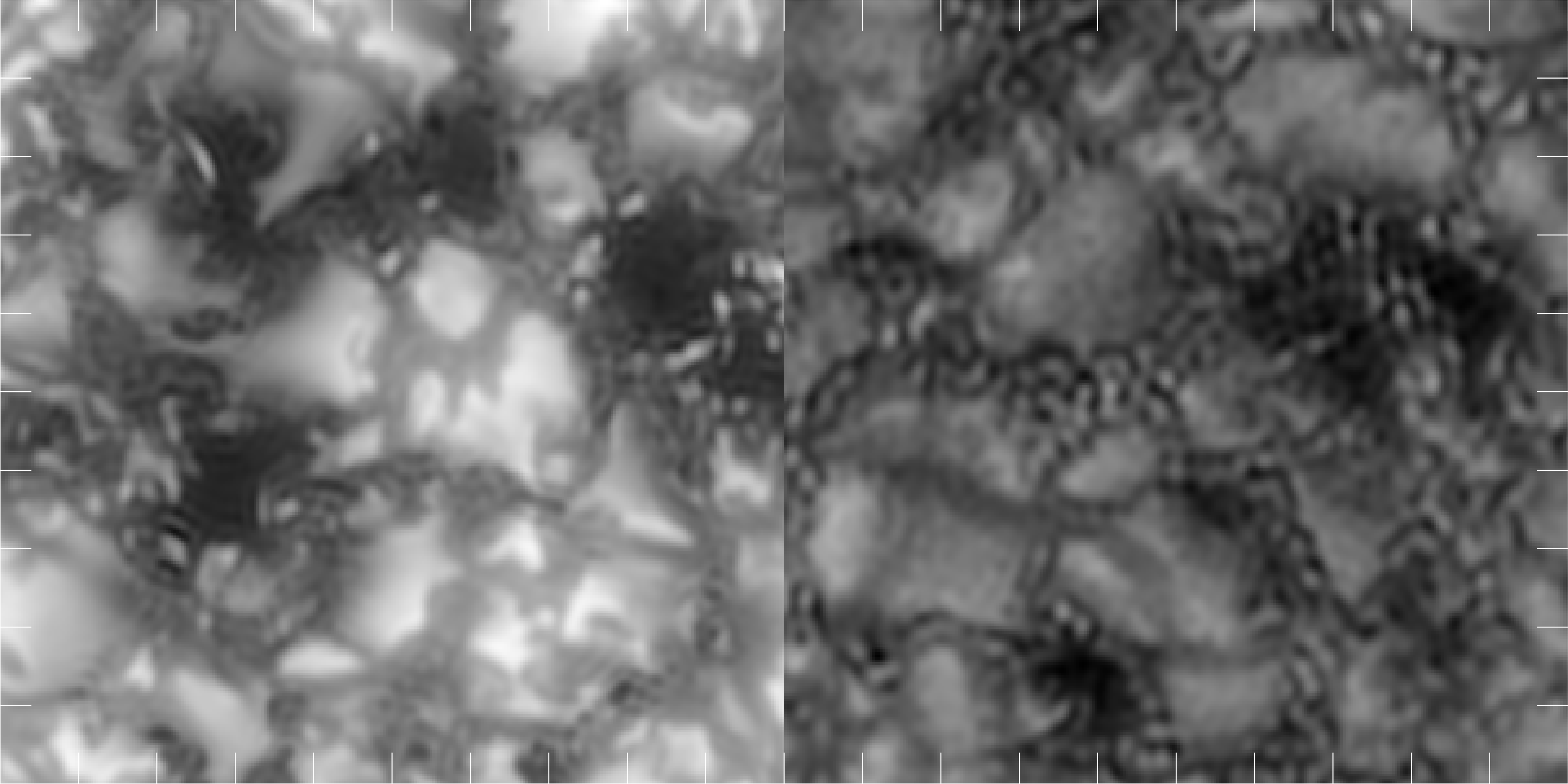}} &  \lab{\centerline{$G_{45}$} 2.39~$\mathrm{g\ cm^{-2}}$} \\ 

 \hspace*{-7mm}
  \lab{\centerline{(7)} 3.56~$\mathrm{g\ cm^{-2}}$}  & \resizebox{!}{4.0cm}{ \includegraphics{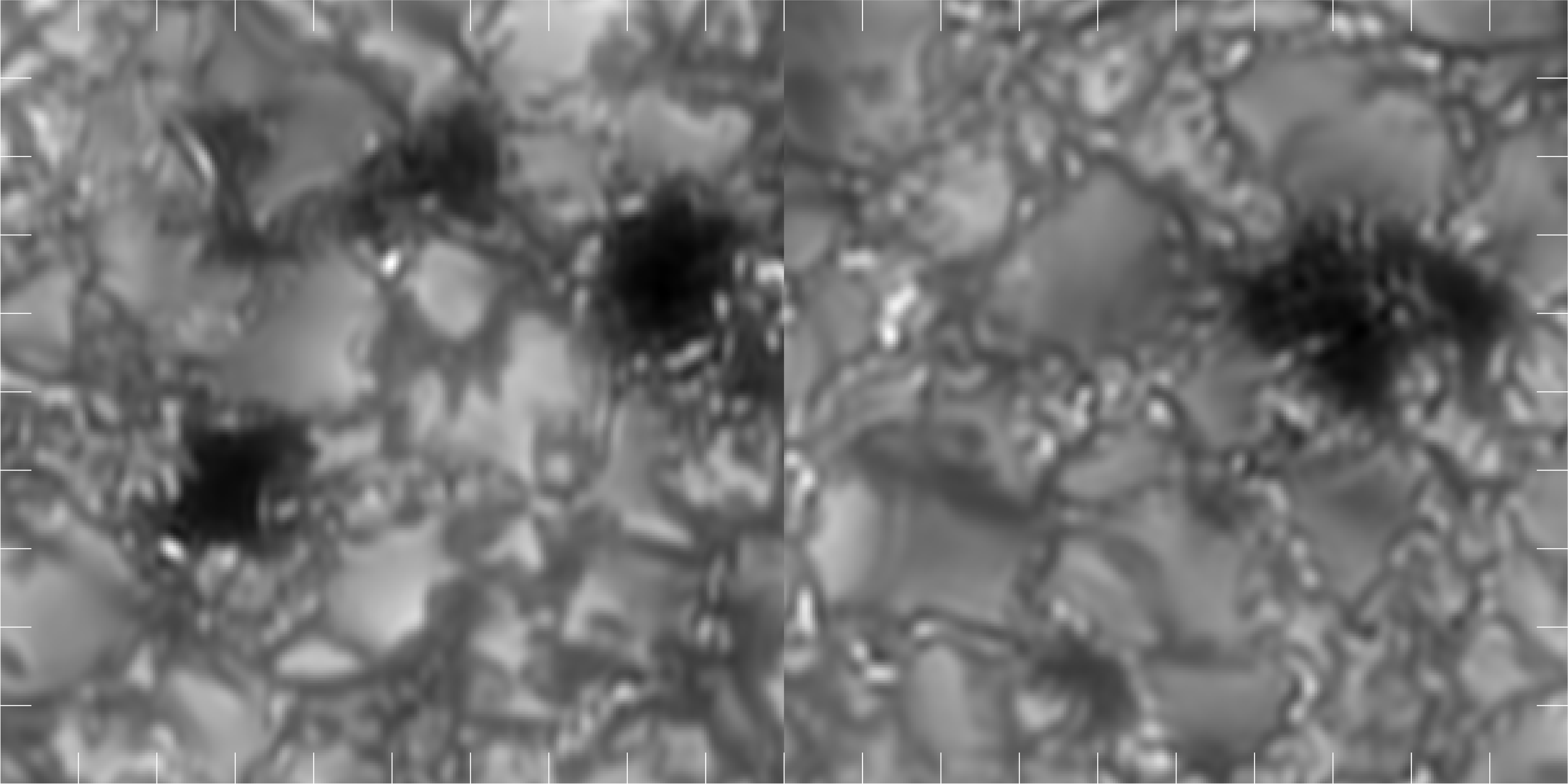}}   &   \resizebox{!}{4.0cm}{ \includegraphics{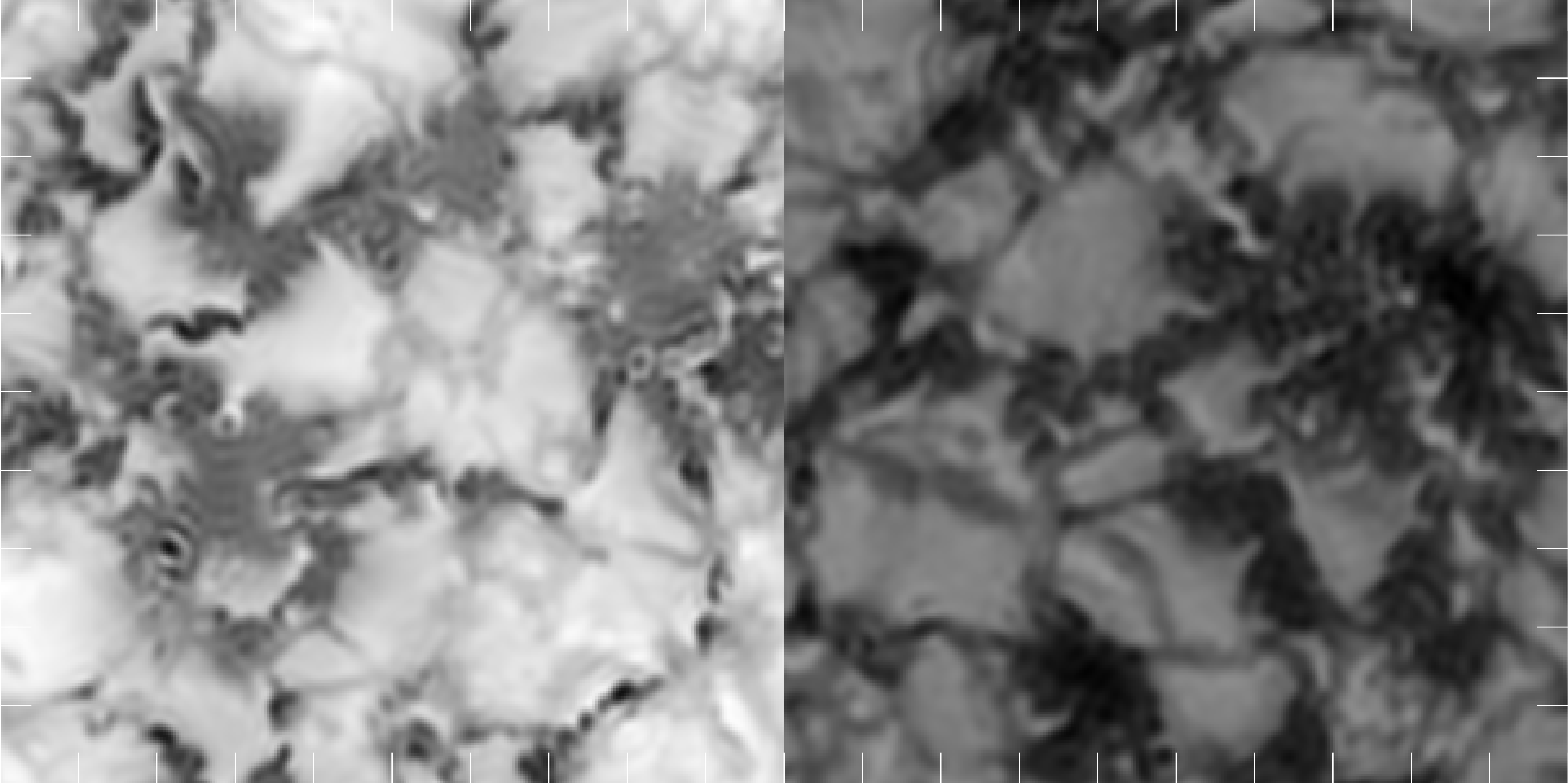}} & \lab{\centerline{$G_{57}$} 3.09~$\mathrm{g\ cm^{-2}}$} 
    
      \end{tabular}
   \caption{\footnotesize An active patch with pores and ribbons of bright points (the pore region close to the vertex of the sunspot in Fig.~\ref{wideband}) compared with the synthetic images from the 3D MHD simulation snapshot from Lagerfj\"{a}rd \& Nordlund (in prep.). From left to right: intensity from synthetics, intensity from observations, gradient maps from synthetics and gradient maps from observations. The depths in terms of column mass shown on the left side are the average depth from all the pixels when computed as in Sect.~\ref{sub:depth}. Each tick-mark corresponds to one arcsec. Each pair of synthetic and observational images are scaled in the same way.}
      \label{tmaps}
      
\end{figure*}

\section{Conclusions}

Temperature stratification is extractable from filtergrams in the \ion{Ca}{ii}~H wings at very high spatial resolution (up to 0\farcs10) and over a wide atmospheric range in the same way as it has been used before in spectrograms. Temperature extraction was possible with a simple analytical procedure making use of a filter with PSFs and passbands that vary with wavelength.

The technique developed in this work to address non-simultaneity of frames in the context of image reconstruction using MOMFBD successfully minimizes spurious signal from the non-simultaneity of the recorded images and is already being adopted for other projects (e.g., \cite{2012ApJ...757...49W,2012ApJ...752..108S,2012A&A...540A..19S}).

The limitations due to the broad and variable filter profile, as discussed in Sect.~\ref{section5}, could be mitigated by introducing an additional rotation stage tilting an interference filter further out into the \ion{Ca}{ii}~H blue wing (for example, the 396.47 nm interference filter mentioned in Sect.~\ref{sect:setup}). This would minimize the tilt angle necessary to scan the whole wing. A more satisfactory solution would be to replace the setup with a tunable filter with narrow passbands and stable PSF (e.g. a Fabry-P\'{e}rot). Significantly narrower passbands would also dramatically increase the sampled height range (see Sect.~\ref{section5}) and allow proper separation of chromospheric and upper photospheric features that currently are mixed in the core filtergrams. 

A way to address the issues with the limitations of the temperature-extraction method used here could be to try to allow some quantities to be height dependent or to iterate temperature in the deep layers as done for spectra by \cite{2002A&A...389.1020R}. However, in filtergrams, the range of the atmosphere to be iterated needs to be much larger. Thus, the ideal solution for this issue is to adapt a fully automatic inversion technique to the \ion{Ca}{ii}~H line wings such as those of  \cite{1997ApJ...478L..45B,1998ApJ...507..470S,2012A&A...543A..34D} and \cite{2000A&A...358.1109F}.

The degree to which simulation data and observations match at several different heights in the \ion{Ca}{ii}~H wing filtergrams is striking. This includes the morphology in differential maps and the general progression of structures with height, excluding the smallest bright points. 

Temperature gradient maps are an interesting way to look at the morphology of structures and a testament to the success of the imaging-processing techniques presented in this work. The large difference in gradient between areas with a strong magnetic field and non-magnetic areas in the outer wings of \ion{Ca}{ii}~H makes these a possible proxy for automatically identifying both magnetic bright and dark features in a single map in deep layers. \cite{2006A&A...452L..15L} explored the blue wing of \ion{Ca}{ii}~H as a proxy magnetometry diagnostic for bright points using differential images from synthetic data at 396.73~nm and 396.48~nm (approximately the $\lambda_{\rm{m},i}$ of position (2) and position (4) respectively). Our maps suggest that it is best to use differentiation blue-ward of that range from the continuum up to 396.43~nm when using this line as a proxy. \cite{2006A&A...452L..15L} also concluded that $\rm{H}\alpha$ is the best proxy. However, the \ion{Ca}{ii}~H wing allows higher resolution. Additionally, it might be more interesting for instruments that target the entire \ion{Ca}{ii}~H line (for example, for its chromospheric potential plus context information) or that focus on observing in the UV. That would be the case, for example, of the upcoming IRIS mission \citep{2011SPD....42.1512L}, which does not include polarimetry.

In our upcoming work, we intend to interpret the temperature stratification obtained for the sunspot in the field of view.

The coefficient $\Gamma_\mathrm{W} = 1.33 \times 10^{-8}$ is calculated by \cite{2002A&A...389.1020R} from the broadening cross-section of \cite{1998MNRAS.300..863B}. 

\begin{acknowledgements}
I am thankful to Dan Kiselman and G\"oran Scharmer for support, supervision, and many discussions. I am indebted to \r{A}ke Nordlund and Anders Lagerfj\"{a}rd for the unpublished MHD simulation snapshot and to Mats Carlsson for sharing the latest MULTI and the HD simulation snapshot. I thank Mats L\"ofdahl for invaluable assistance with the tiltable filter and Peter S\"{u}tterlin for a solution to the PD camera. I am also thankful to Michiel van Noort and Jaime de La Cruz Rodr\'{i}guez for general assistance.
This project was supported by a Marie Curie Early Stage Research Training Fellowship of the European Community’s Sixth Framework Programme under contract number  MEST-CT-2005-020395: The USO-SP International School for Solar Physics and by the H.M. Konungens Wallenberg fond.
The Swedish 1-m Solar Telescope is operated on La Palma by the Institute for Solar Physics of the Royal Swedish Academy of Sciences in the Spanish Observatorio del Roque de los Muchachos of the Instituto de Astrof\'isica de Canarias. 
\end{acknowledgements}


\end{document}